\documentclass[12pt]{article}
\usepackage[english]{babel}
\usepackage[a4paper,top=2cm,bottom=2cm,left=2cm,right=2cm,marginparwidth=1.75cm]{geometry}
\usepackage{authblk}
\usepackage{graphicx}
\usepackage{siunitx}
\usepackage{bm}
\usepackage{newtxtext}
\usepackage{newtxmath}
\usepackage[square,sort,comma,numbers]{natbib}
\usepackage{hyperref}
\usepackage{booktabs}
\usepackage{subcaption}
\usepackage{multirow}
\usepackage{epstopdf}
\usepackage{caption}
\usepackage{url}
\usepackage[export]{adjustbox}
\usepackage{wasysym}
\usepackage{float}
\usepackage{soul}
\usepackage[prefix]{nomencl}
\usepackage{doi}
\makenomenclature

\renewcommand{\nomgroup}[1]{%

  \ifthenelse{\equal{#1}{A}}{\item[\textbf{Latin Symbols}]}{%
  \ifthenelse{\equal{#1}{B}}{\item[\textbf{Greek Symbols}]}{%
  \ifthenelse{\equal{#1}{C}}{\item[\textbf{Subscripts}]}{}}}}
\hypersetup{colorlinks = true,linkcolor = blue,anchorcolor =red,citecolor = blue,filecolor = red,urlcolor = red, pdfauthor=author}

\newcommand{\RomanNumeralCaps}[1]
\linenumbers

\DeclareSIUnit\bar{bar}

\title{Bubble size distribution and electrode coverage at porous nickel electrodes in a novel 3-electrode flow-through cell}
\author[1]{Hannes Rox\footnote{E-mail: h.rox@hzdr.de}}
\author[1,2,3]{Aleksandr Bashkatov}
\author[1]{Xuegeng Yang}
\author[4]{Stefan Loos}
\author[1]{Gerd Mutschke}
\author[1]{Gunter Gerbeth}
\author[1,2,3]{Kerstin Eckert\footnote{E-mail: k.eckert@hzdr.de}}

\affil[1]{Institute of Fluid Dynamics, Helmholtz-Zentrum Dresden-Rossendorf, Bautzner Landstrasse 400, Dresden, 01328, Germany}
\affil[2]{Hydrogen Lab, School of Engineering, Technische Universit{\"a}t Dresden, Dresden, 01062 Germany}
\affil[3]{Institute of Process Engineering and Environmental Technology, Technische Universit{\"a}t Dresden, Dresden, 01062, Germany}
\affil[4]{Fraunhofer Institute for Manufacturing Technology and Advanced Materials IFAM, Branch Lab Dresden, Winterbergstr. 28, 01277 Dresden, Germany}

\date{}

\begin{document}

\maketitle

\vspace{-1cm} 
\begin{abstract}
A novel 3-electrode cell type is introduced to run parametrical studies of H$_2$ evolution in an alkaline electrolyte on porous electrodes. 
Electrochemical methods combined with a high-speed optical measurement system \textcolor{black}{are applied simultaneously to characterize the electrodes and the bubble dynamics} in terms of bubble size distribution and coverage of the working electrode. 
    Three different cathodes made of expanded nickel are investigated
    at applied current densities of $|j| = 10$ to  \SI{200}{\milli\ampere\per\centi\metre\squared} without forced flow and at a flow rate of \SI{5}{\milli\litre\per\minute}. The applied current density is found to significantly influence both the size of detached bubbles and the surface coverage of the working electrode. The forced flow through the cathodes is found to strongly reduce
    the bubble size up to current densities of about 100
    mA cm$^{-2}$, whereas the initial transient until the cathode surface is completely covered by bubbles is only marginally affected by the flow-through. 
\end{abstract}

\section{Introduction}
\label{sec:intro}
Water electrolysis using solar- or wind-derived electricity to produce high-purity hydrogen gas is a promising pathway towards a net-zero-emissions industry \citep{Renssen2020, Spek2022}. Hydrogen could replace the fossil fuels often used in industry \citep{Staffell2019} and be an alternative in applications that are hard to directly electrify, e.g., aviation, shipping and inter-seasonal energy storage, or used to produce synthetic natural gas and various synthetic liquid fuels and e-fuels \cite{smolinka2021electrochemical}.
Although today’s production of pure hydrogen gas almost entirely relies on fossil fuels \cite{Spek2022}, the priority is to shift the existing hydrogen production sector toward renewable/green hydrogen using wind and solar energy \cite{Milani2020, Sazali2020, EUstrategy}. \textcolor{black}{Alkaline water electrolysis is still the most mature technology that requires only little of platinum-group metals, although suffering from lower efficiencies and current densities when compared to other technologies.}
One reason for this, among others, is that considerable losses are caused by the generated hydrogen and oxygen bubbles, which  increase the ohmic contribution to the overall overpotential as they block the electrode surface and increase electrolyte resistance \cite{Angulo2020, zhao2019gas, leistra1987voltage, lake2022impact}. To reduce energy losses and thus make green hydrogen cheaper, better knowledge of how to further advance the bubble departure is necessary. 

Besides the operating expenditures (OPEX), which in the long term are mainly caused by the electricity consumption and lack of efficiency, capital expenditures (CAPEX) also need to be decreased \citep{Esposito2017}. One promising approach for a \textcolor{black}{cheaper} new cell design is so-called membraneless electrolyzers. 
In general there are three different types of membraneless electrolyzers: flow-through \citep{Gillespie2015, Gillespie2017, Gillespie2018, Bui2020, Davis2019, Neil2016}, flow-by \citep{Pang2020, Hashemi2015, Hashemi2019} and decoupled electrolyzers \citep{Dotan2019,Yan2021,Solovey2018a,Solovey2018b,Solovey2021}. Here, we focus on flow-through electrolyzers, where the electrolyte flowing through \textcolor{black}{porous} electrodes is used to keep the products H$_2$ and O$_2$ separated by directly flushing them out of the cell in separate channels \citep{Gillespie2015, Gillespie2017, Gillespie2018, Bui2020, Davis2019, Neil2016}. 
\textcolor{black}{The elimination of the membrane in flow-through electrolyzers both simplifies the overall design and improves the impurity tolerance, thus enabling the operation with tap water \citep{Esposito2017}. As a result, a significant reduction of the CAPEX and OPEX can be expected. Hence, with decreasing costs for electrical power, the produced hydrogen could become economically competitive \citep{Esposito2017}.}

The diameter of the bubbles $d_\text{B}$ at the detachment and the coverage of the electrode $A_\text{cov}$ by those bubbles are critical parameters for effectively separating the products using the electrolyte flow and for achieving satisfactory overall efficiency \citep{Angulo2020, Gillespie2015, Davis2019}.
In membraneless water electrolyzers, the efficiency is already reduced by higher ohmic losses compared to zero-gap alkaline or polymer electrolyte membrane (PEM) electrolyzers \citep{Esposito2017}. Therefore, it is crucial to optimize the geometry and surface of the porous electrodes of the membraneless design.
\citeauthor{Gillespie2015} \citep{Gillespie2015} performed a detailed study on the relation between the electrode gap, flow velocity and applied current density for a membraneless divergent-electrode-flow-through (DEFT) cell. 
Higher electrode coverage and the possible blocking of pores lead to higher losses due to the increasing pressure drop and the higher ohmic resistance. More importantly, bubbles at the inner site of each electrode can form a gas meniscus, resulting in the crossover of the products as soon as they overlap \citep{Gillespie2015}. \textcolor{black}{Similar effects of the electrode gap were reported by \citeauthor{Pang2020} \citep{Pang2020} and \citeauthor{Rajaei2021} \citep{Rajaei2021}.}

The electrogenerated gas bubbles growing at the electrodes experience a number of forces including electric, hydrodynamic, thermocapillary forces, and buoyancy \citep{Yang2018, Bashkatov2019, Hossain2020, Meulenbroek2021, hossain2022, bashkatov2022}.
When the bubbles are surface attached, they experience \textcolor{black}{additional contact pressure and surface tension forces} that needs to be further overcome prior to the bubble departure, leading to an increase in the overpotential \citep{Zhang2010}. Furthermore, bubble-bubble coalescence events are known to promote faster detachment, especially at a high current density $j$.  However, the bubbles growing inside a porous electrode are generally more easily entrapped due to densely packed catalytic surface areas and weak convective flows.

In general, porous electrodes such as meshes or foams offer large areas for electrochemical reactions, and there are
many cavities on the surface, e.g., in the weave knots of woven meshes. This leads to a higher number of nucleation sites compared to planar electrodes \citep{Li2021c}. If the geometry or the surface of the electrodes are adjusted, the bubble dynamics can be improved such that smaller bubbles detach faster. In general, small pore sizes lead to a homogeneous flow distribution and a high reactive surface area \citep{Rajaei2021}. However, the bubbles entrapped inside the electrode lead to an increase in the overpotential and in the pressure drop \citep{Zhang2010}. \citeauthor{Lee2021b} \citep{Lee2021b} performed a study on the structural effect of an electrode mesh. For the expanded meshes, it was proven that the overpotential \textcolor{black}{decreases} as the ratio of pore to strand width converges to 1. From \textcolor{black}{this} study it can be concluded that a hydrophilic surface will favor a re-wetting of the electrode and, by that means, remove gas bubbles with a smaller departure diameter and re-cover the active area. \textcolor{black}{A similar influence of the mesh structure on the efficiency was also reported by \citeauthor{Zhang2010} \citep{Zhang2010}.}

Structures incorporated into the porous electrode, such as tapered or expanding channels, can impose a capillary force which enhances the bubble detachment and the bubble transport inside the porous electrode \citep{Angulo2020, Reyssat2014}. Another possibility is using different pore sizes in the inner and outer regions of a porous electrode. This can stimulate bubble coalescence and splitting processes in a targeted manner \citep{Yang2022}. Besides the bubble detachment, the efficiency of the gas evolution has to be taken into account. This is affected by the competing processes of mass transfer to the liquid bulk and to the adhering bubbles \citep{Vogt2011}. The mass transfer is influenced by the bubble coverage, which moreover affects the actual current density  \citep{Vogt2011, Eigeldinger2000, Vogt2015, Vogt2017}.

The present study uses a novel, 3D-printed membraneless electrolysis cell developed as a platform for analyzing the bubble dynamics and electrochemical performance of porous electrodes. \textcolor{black}{By simultaneously characterizing the size of the detached bubbles, $d_\text{B}$, the electrode coverage $A_\text{cov}$ and overpotential losses over a wide range of current densities $j$, it is shown that the current density influences both, $d_\text{B}$ and $A_\text{cov}$, significantly. Additionally, the effect of the applied electrolyte flow through the porous electrode is studied.}
This new type of cell provides a  
reproducible and uniform way to jointly characterize both the electrochemical performance and the bubble dynamics in terms of bubble size distribution and electrode coverage for different electrode geometries, materials or coatings over a wide range of parameters.
\textcolor{black}{Thus, the new cell has  the potential to become a valuable tool in the simultaneous optimization of the electrochemical properties and the resulting bubble dynamics of new electrode materials, geometries and coatings.}

\section{Materials and Methods}
\label{sec:methods}
\subsection{Cell design}
\label{sec:cell}
\begin{figure}[ht]
\centering
    \includegraphics[width=0.7\textwidth]{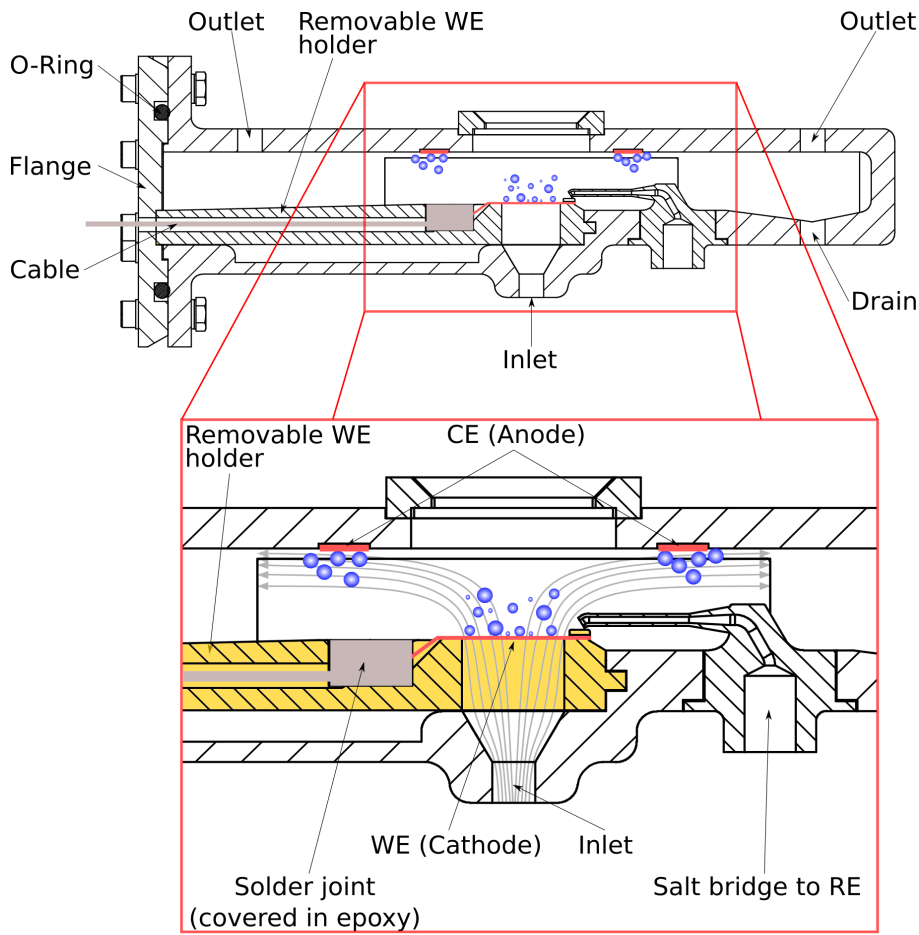}
	\caption{Schematic of the electrochemical cell consisting of three electrodes and a detailed view of the examined WE with indicated streamlines of the electrolyte flow. Note the two observation windows, \textcolor{black}{detailed in} Fig. \ref{fig:scheme_setup}, and the removable WE holder (highlighted in yellow). \label{fig:scheme_cell}}
\end{figure}
\begin{figure}[ht]
\centering
	\includegraphics[width=\textwidth]{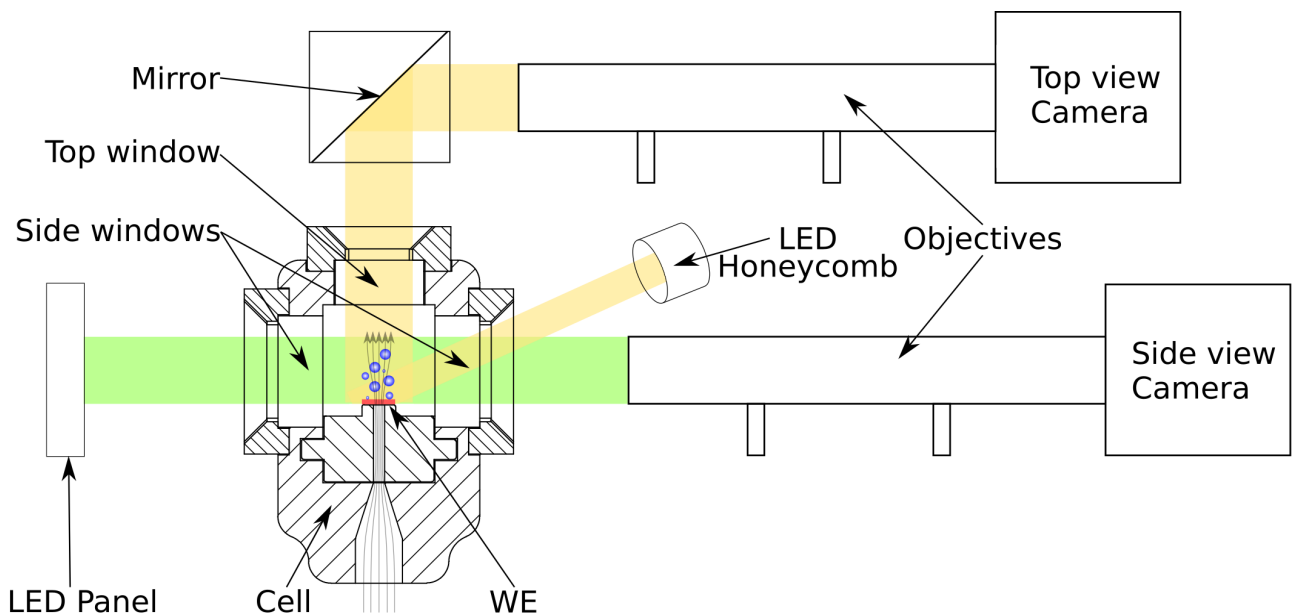}
	\caption{Schematics of the optical measurement system for both the top and side view images of the WE \label{fig:scheme_setup} (colors are for illustration only)}
\end{figure}

\newpage
Fig. \ref{fig:scheme_cell} documents the schematic design of the 3D printed electrochemical cell. The bottom part of Fig. \ref{fig:scheme_cell} is a magnified view of the electrodes comprising part of the cell.
The electrochemical cell consists of a horizontally installed cathode (working electrode, WE) and two anodes (counter electrodes, CEs), as well as of a reversible hydrogen electrode (RHE, HydroFlex\,\textsuperscript{\tiny\textregistered}, Gaskatel, Germany) serving as a reference electrode (RE)
connected to the cell via a salt bridge. 
The cathode is a section of expanded metal (EM) nickel of approx. 1$\times$\SI{10}{\milli\metre\squared} (see Section \ref{sec:fabrication}) and the anode consists in two pieces of Pt foil (purity 99.95~\%) together having an area of $\approx \SI{100}{\milli\metre\squared}$.
The electrochemical cell features two observation windows and an easily exchangeable cathode holder \citep{Bashkatov2022FC3}. It is worth emphasizing the possibility to simultaneously record side view and top view images of the hydrogen-forming cathode to capture the bubble evolution over time from two different perspectives (see Fig. \ref{fig:scheme_setup}). To allow the top view, the anode electrode is produced as two pieces of foil at a distance from each other as shown in Fig. \ref{fig:scheme_cell}. The details of the electrochemical system are given in Section \ref{sec:electrochemical_methods}.

The cell and the electrode holders were printed from \textcolor{black}{DraftGrey (KOH resistant, Stratasys, USA)} using an Objet30 Prime V5 (Stratasys, USA). The observation windows, made of PMMA (Poly(methyl methacrylate)), were glued into the cell using epoxy adhesive (Araldite 2013-1, Huntsman, USA). The configuration of the cell is T-shaped and therefore possesses one electrolyte flow inlet, located below the cathode, and two outlets located as shown in Fig. \ref{fig:scheme_setup}. 
The inlet and outlets are connected to separate reservoirs to avoid electrolyte remixing. To achieve a constant electrolyte flow without any pulsations, a microfluidic controller was used of the type OB1 MK3+ (Elveflow, France) with a maximum pressure of \SI{2}{\bar}. Within the scope of the present study, the possible mixing of H$_2$ and O$_2$ gases is tolerated, as the main focus is on the bubble dynamics at the cathode electrode of a different geometry. Therefore, no further precautions are taken to separate the products at the anode and cathode.

\subsection{Fabricating the working electrode}
\label{sec:fabrication}
\begin{figure}[h]
\centering
	\includegraphics[width=0.5\textwidth, frame]{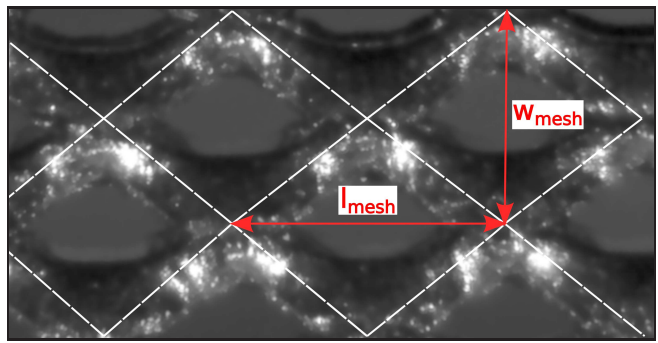}
	\caption{Image of one expanded nickel metal piece with diamond-shaped pores and the two essential parameters required to differentiate the electrodes: the mesh width $w_\text{mesh}$ and the mesh length $l_\text{mesh}$}
	\label{fig:example_EM}
\end{figure}

Sections of expanded metal nickel (Benmetal, Germany) were cut with a laser into pieces of 1 $\times$ \SI{20}{\milli\metre\squared}. 
The cut electrodes were soldered to a copper wire ($d = \SI{1}{\milli\metre}$) and glued to a 3D printed electrode holder using epoxy adhesive. The free surface area of the electrode was limited to approx. 1 $\times$ \SI{10}{\milli\metre\squared}
by covering the rest of the electrode area and the solder joint with epoxy. Afterwards, the electrodes were cleaned using deionized water and ethanol. To characterize the electrode surface, three nickel foils (GoodFellow, purity 99.99 \%) were produced as benchmark electrodes in the same procedure, except that the surfaces of these foils were polished using \SI{1}{\micro\metre} diamond polish and alumina polish (PK-4 Polishing Kit, BASi, USA) before the cleaning procedure. All the characteristic parameters of the expanded metal sections were determined by Benmetal using an OSIF MeshInspector ML system. In addition, the hydraulic diameter \textcolor{black}{$d_\text{h}$} was calculated using the following equation and assuming the pores to be diamond-shaped (see Fig. \ref{fig:example_EM} and Table \ref{tab:electrodes}). For this purpose, the length, $l_\text{pore}$, and width, $w_\text{pore}$, of the mesh opening provided by Benmetal (see Supplemental Material) were used  to calculate the area and perimeter of the pore ($A_\text{pore}$ and $P_\text{pore}$). \textcolor{black}{In the following, the electrodes are named according to the mesh width $w_\text{mesh}$, which is defined as the distance between the centres of two junctions in the direction of the short diagonal (see Fig. \ref{fig:example_EM}).}

\begin{equation}
    d_\text{h} = \frac{4 \cdot A_\text{pore}}{P_\text{pore}} = \frac{w_\text{pore}\cdot l_\text{pore}}{\sqrt{w_\text{pore}^2 + l_\text{pore}^2}}
    \label{eq:hydraulic_diameter}    
\end{equation}

\begin{table}[ht]
   \centering
    \caption{Overview of the mesh width and length ($w_\text{mesh}$ and  $l_\text{mesh}$), the calculated hydraulic diameter $d_\text{h}$ (see Eq. \ref{eq:hydraulic_diameter}), the electrode porosity $\varepsilon$ and electrode thickness $t_\text{el}$ and the determined electrochemical characteristics, double-layer capacitance $C_\text{D}$ and electrochemically active surface area $ECSA$ of the EM and the Ni benchmark electrodes Ni\_bm}
    \begin{tabular}{c||c c c c c |c c}
        \toprule
        \multirow{2}{*}{\textbf{Name}} & \boldmath$w_\text{mesh}$ & \boldmath$l_\text{mesh}$ & \boldmath$d_\text{h}$ & \boldmath$\varepsilon$ & \boldmath$t_\text{el}$ & \boldmath$C_\text{D}$ & \boldmath$ECSA$ \\
        & in \si{\micro\metre} & in \si{\micro\metre} & in \si{\micro\metre} & in \% & in \si{\micro\metre} & in \si{\micro\farad} & in \si{\centi\metre\squared}\\
        \midrule
        EM\_475 & 475 & 600 & 136.13 & 22.6 & 75 & 12.1721 & 0.3453 \\
        EM\_500 & 500 & 600 & 86.10 & 11.8 & 150 & 10.6417 & 0.3019 \\
        EM\_518 & 518 & 602 & 100.86 & 15.5 & 200 & 17.6738 & 0.5014 \\
        Ni\_bm & - & - & - & -  & 100 & 8.4916 & 0.2409 \\
        \bottomrule
    \end{tabular}
    \label{tab:electrodes}
\end{table}

\subsection{Electrochemical methods}
\label{sec:electrochemical_methods}
The electrochemical experiments were carried out in 1 M KOH (Titripur, Merck, Germany) at \SI{293}{\kelvin} in the three-electrode cell described above (see Section \ref{sec:cell}). 
The RE was placed in a separate syringe and connected to the cell through a salt bridge consisting of a 3D printed capillary and a tube ($d_\text{in} = \SI{0.5}{\milli\metre}$). All WE potentials were measured with respect to the RE using an electrochemical workstation (CHI660E, CH Instruments, USA).

To estimate the onset potential $E_\text{O}$ of the hydrogen evolution reaction (HER) for all WEs, Linear Sweep Voltammetry (LSV) was performed over a potential range from \SI{0}{\volt} to \SI{-0.8}{\volt} at a scan rate of $\nu =  \SI{50}{\milli\volt\per\second}$.
The onset potential $E_\text{O}$ is defined as the intersection of the fitted tangent with the zero axis. Cyclic Voltammetry (CV) was used to calculate the double-layer capacitance $C_\text{D}$ based on the following equation 
\begin{equation}
    C_\text{D} = \frac{I_\text{a} + |I_\text{c}|}{2 \cdot \nu},
\end{equation} 
where $I_\text{a}$ is the absolute anodic current and $I_\text{c}$ the cathodic current. The electrochemically active surface area $ECSA$ was determined by multiplying the calculated capacitance $C_\text{D}$ with the ratio between $ECSA_\text{bm}$ and $C_\text{D, bm}$ of the smooth benchmark surface \citep{Skibinska2022}. This is defined by the averaged surface of the three benchmark electrodes made out of plain nickel foil (see Section \ref{sec:fabrication}).
\begin{equation}
    ECSA = C_\text{D} \cdot \frac{ECSA_\text{bm}}{C_\text{D, bm}}
    \label{eq:ecsa}    
\end{equation}
The measured electrochemically active surface area $ECSA$ was used to calculate the current $I$ \textcolor{black}{to be applied for achieving} a specific current density $j$ ($I = j \cdot ECSA$). For the imaging of the bubble nucleation, growth and detachment, galvanostatic measurements were performed at a sample rate of \SI{50}{\hertz} with various current densities ($j = -10, -20, -50, -100$ or $  \SI{-200}{\milli\ampere\per\centi\metre\squared}$). 

\subsubsection*{Experimental parameters}
For \textcolor{black}{each} WE, all measurements, including the electrochemical characterization, were performed under normal conditions (approx. $T = \SI{293}{\kelvin}, p = \SI{1}{\bar})$ \textcolor{black}{within the same day} to avoid any contamination. Table \ref{tab:expparameters} gives an overview of all experimental parameters. For each parameter set, three measurements were performed in order to obtain statistical confidence. Thus, a total of $3 \times 5000$ images were used for the side view evaluation and $3 \times 1000$ for the top view evaluation. Fewer images were used for the top view because the rising bubbles \textcolor{black}{tended to cover} the observation window over time. The galvanostatic measurements were performed over a period of \SI{20}{\second}.
\begin{table}[ht]
    \centering
    \caption{Overview of the varied parameters, the applied current density $j$, electrolyte flow rate $\dot{V}$, electrochemically active surface area $ECSA$ and electrode porosity $\varepsilon$, as well as the constant experimental parameters, the electrolyte concentration $c_\text{KOH}$ used and the free surface of the Pt-CE $A_\text{CE}$}
    \begin{tabular}{c|cl}
        \toprule
        & \textbf{Parameter} & \textbf{Description} \\
        \midrule
        \multirow{4}{*}{\rotatebox[origin=c]{90}{\parbox[c]{2cm}{\centering Varying}}} & $j$ & -10, -20, -50, -100, \SI{-200}{\milli\ampere\per\centi\metre\squared} \\
        & $\dot{V}$ & 0 and \SI{5}{\milli\litre\per\minute} \\
        & $ECSA$ & 30.19, 34.53, \SI{50.14}{\milli\metre\squared} (see Table \ref{tab:electrodes}) \\
        & $\varepsilon$ & 11.8, 15.5, 22.6 \% (see Table \ref{tab:electrodes}) \\
        \midrule
        \multirow{2}{*}{\rotatebox[origin=c]{90}{\parbox[c]{1cm}{\centering Const}}} & $c_\text{KOH}$ & 1 M KOH \\
        & $A_\text{CE}$ & \SI{80}{\milli\metre\squared} \\
        \bottomrule
    \end{tabular}
    \label{tab:expparameters}
\end{table}

\subsection{Imaging and image processing}
Two high-speed cameras (IDT OS-7 S3, USA), each equipped with a precision micro-imaging lens with a magnification of 2 (Optem\textregistered \, FUSION, USA), were used at a sample rate of \SI{500}{fps} and a bit depth of \SI{12}{bit}. 
The spatial resolution of the side view and top view cameras was \SI{593}{px\per\milli\metre} and \SI{505}{px\per\milli\metre}, respectively.
Both calibrations were performed using a dual-axis linear scale micrometer with a scale division of \SI{25}{\micro\metre} (Edmund Optics, USA). Greater details on the camera settings in the form of metadata can be found in the Supplemental Material (.hdf5 files). To complete the shadowgraphy system, one LED panel (CCS TH2, Japan) and two honeycomb LEDs (IDT, USA) were incorporated, providing back and top illumination, respectively.
The side view images were used to calculate the size of the bubbles at the detachment, whereas the top view images made it possible to estimate the electrode coverage over time.
The image analysis procedures were performed in Python 3.8. 
\begin{figure}[ht]
    \centering
	\includegraphics[width=0.7\textwidth]{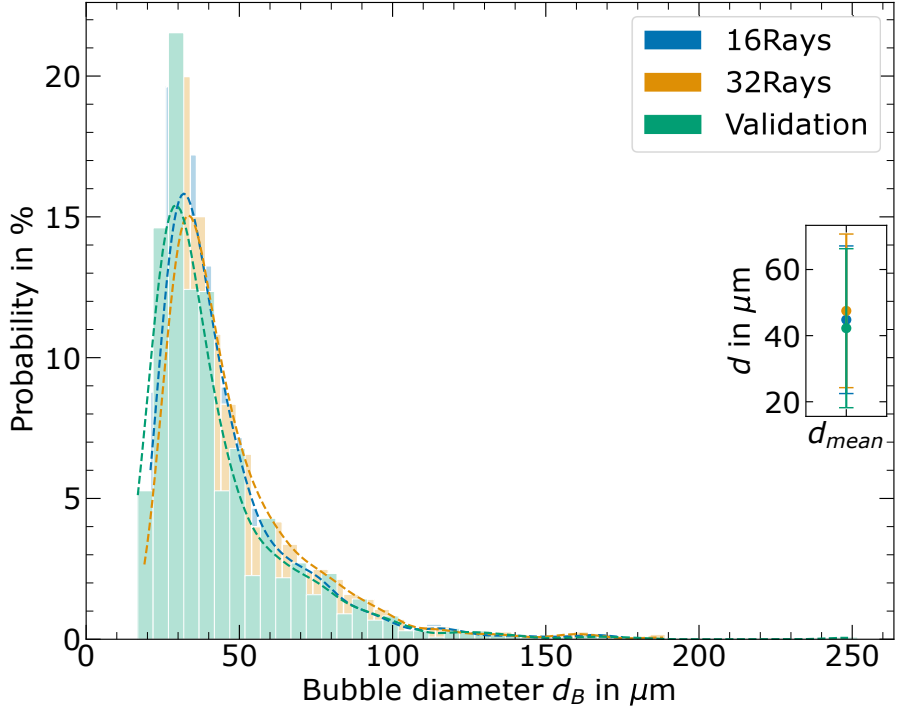}
	\caption{Bubble size distribution with dashed fit for better comparison and mean bubble diameter (plotted in inset axis) for two different stardist models (16 rays (blue) and 32 rays (orange)) in comparison to the validation data set (green) with semi-automatically annotated bubbles showing a good agreement of both models \label{fig:validation_stardist}}
\end{figure}
\begin{figure}[ht]
     \centering
     \begin{minipage}{0.48\textwidth}
        \begin{subfigure}[t]{\textwidth}
            \centering
            \includegraphics[width=\textwidth, frame]{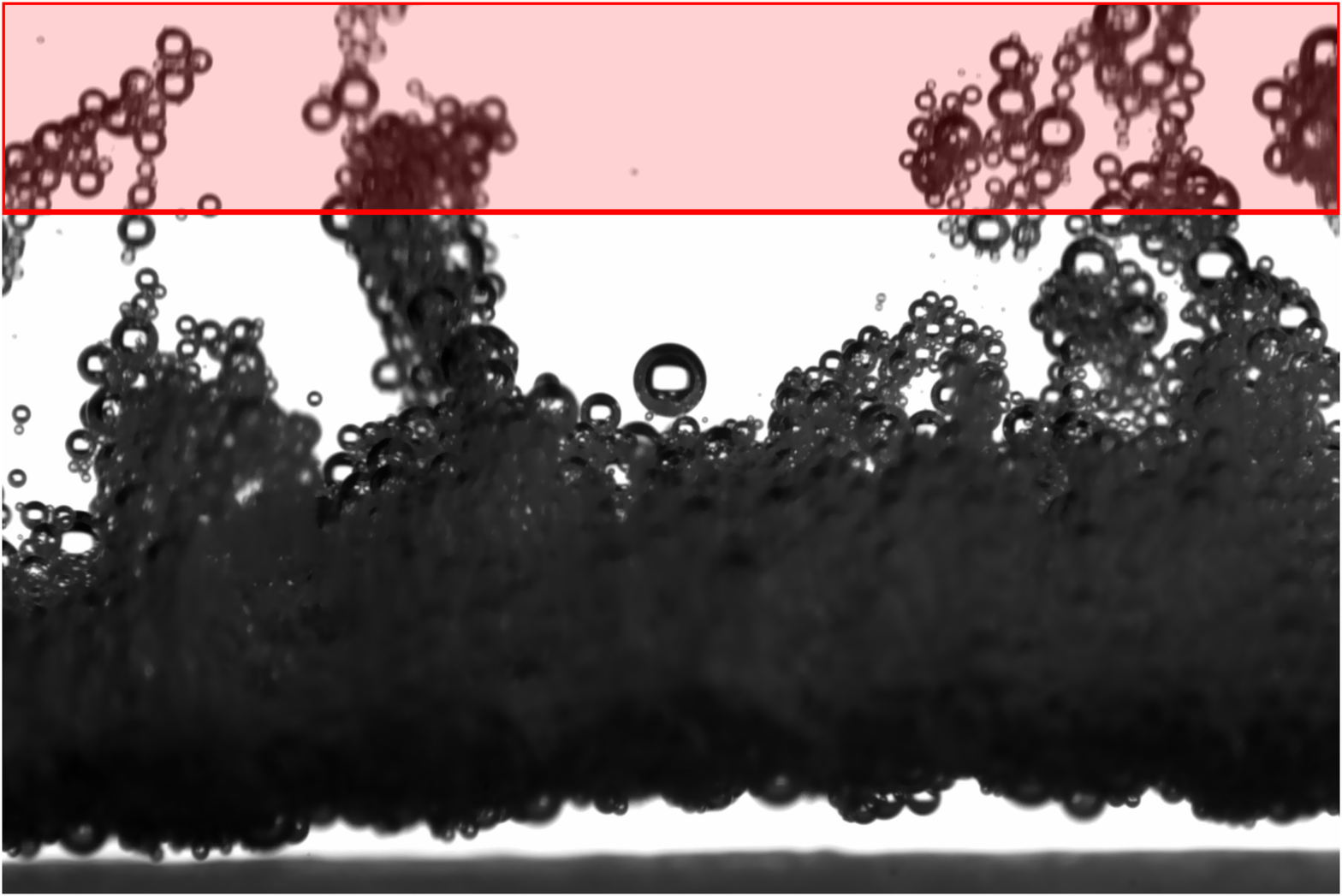}
            \caption{Raw image of side view analysis}
            \label{fig:sideview_raw}
        \end{subfigure}
     \end{minipage}
     \hfill
     \begin{minipage}{0.48\textwidth}
         \centering
         \begin{subfigure}[t]{\textwidth}
            \centering
            \includegraphics[width=\textwidth, frame]{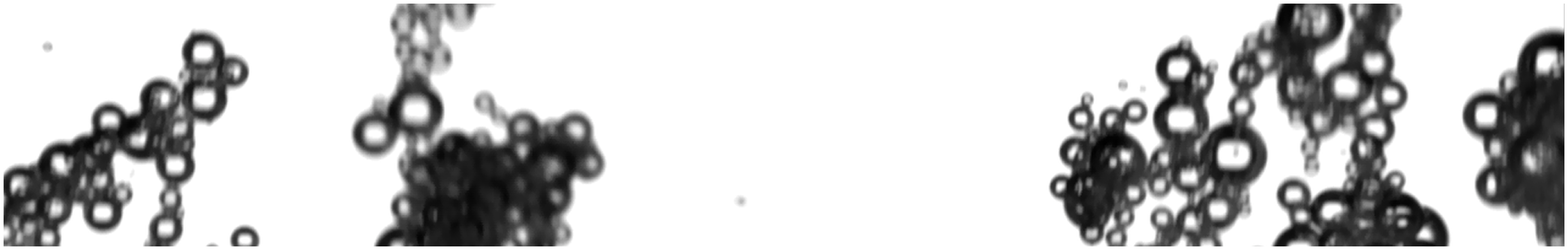}
            \caption{Cropped ROI}
         \end{subfigure}
         \vfill
         \begin{subfigure}[b]{\textwidth}
            \centering
            \includegraphics[width=\textwidth, frame]{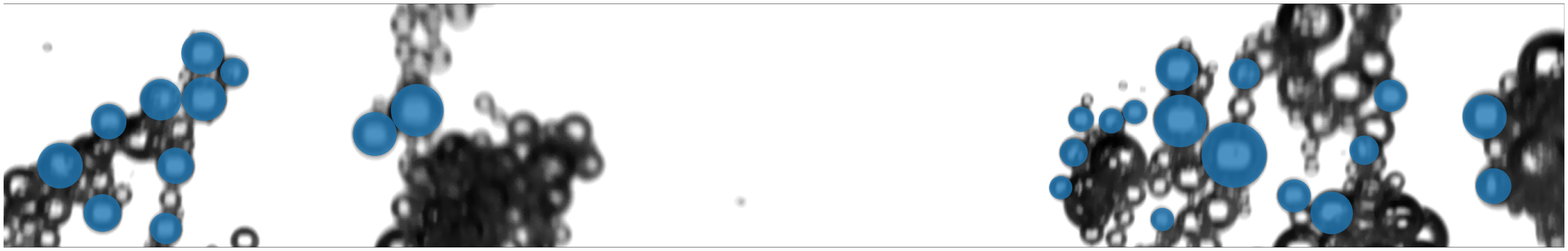}
            \caption{Result with annotated bubbles for all bubbles detected using stardist}
         \end{subfigure}
     \end{minipage}
     \caption{Processing of side view images consisting of cropping the ROI, detecting the bubbles using stardist and post-processing (deleting blurred bubbles and misdetected objects)}
     \label{fig:image_processing_sideview}
\end{figure}

The detection of single bubbles using conventional, well-established algorithms such as the Hough transform becomes especially challenging at high current densities, where a high fraction of gas bubbles overlap. Therefore, an approach based on machine learning was applied using stardist \citep{Schmidt2018, Weigert2020}.
The stardist method is based on the prediction of the object probability $d_\text{i,j}$ and the Euclidean distance $r_\text{i,j}^\text{k}$ along $k$ radial rays. Each bubble is described by multiple rays as a star-convex polygon. These were fitted as an ellipse to calculate the bubble diameters $d_\text{B}$ \citep{Hessenkemper2022}. 
Since the key criterion in the present analysis is the precise detection of the bubble size distribution, the bubble size $d_\text{B}$ is used as a criterion for evaluating the model. Two different stardist models, 16 or 32 rays, respectively, were compared to an evaluation set of randomly chosen set of 100 images (see Supplemental Material). A comparison of the results (see Fig. \ref{fig:validation_stardist}) led to the choice of the model using 16 rays, as the deviation is minimal. Therefore, only the results calculated with the 16-ray model will be discussed in the following. Since the evolving H$_2$ bubbles can be described as spheres, the number of 16 points from the rays at the bubble boundary is also enough for the ellipse fitting algorithm.

\begin{figure}[h]
    \centering
	\includegraphics[width=0.55\textwidth]{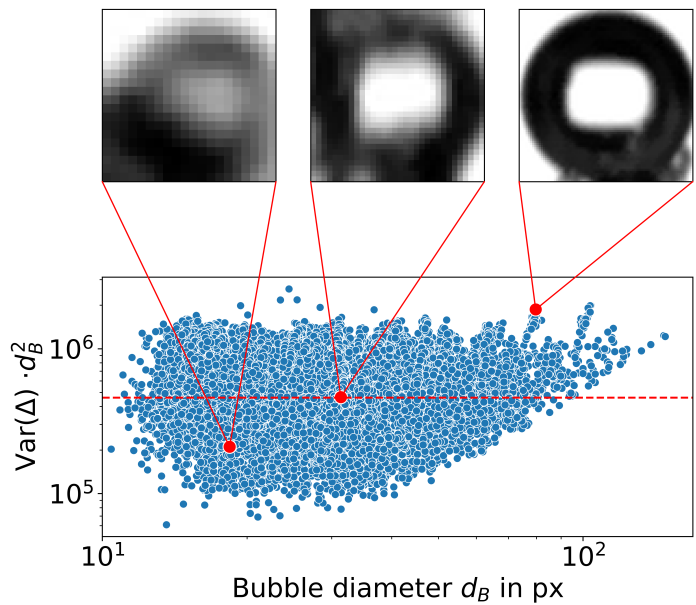}
	\caption{Blur criterion (dashed red line) to distinguish between blurred and sharp bubbles with example images of a blurred and a sharp bubble, as well as a bubble in the transition area \label{fig:DOF_criteria}}
\end{figure}

Another advantage of stardist is that less image preprocessing is needed. Only the region of interest (ROI) needs to be cropped out of the image and the bubbles are detected using the trained model. As stacks of bubbles form at the electrode (see Fig. \ref{fig:sideview_raw}), a rectangular region with a width of \SI{30}{px} at the top of the images was chosen as the ROI to ensure that only detached bubbles are processed. The postprocessing is split into deleting blurred bubbles and linking all those remaining (see Fig. \ref{fig:image_processing_sideview}b-c). 
The phenomenon underlying blurred bubbles is the depth of field (DOF) of the optical system. By moving the calibration plate through the plane of focus, the DOF could be estimated as $\approx \SI{66}{\micro\metre}$ (see  Supplemental
Material). The variance in the Laplacian of the single bubble was chosen as the criterion to define a bubble inside the measurement plane. Since this value is dependent on the bubble size $d_\text{B}$ or the size of the image section, respectively, a threshold is applied first for deleting misdetected objects ($Var(\Delta) < 50$). The variance is then multiplied by the bubble diameter $d_\text{B}$ to guarantee that all bubble sizes are treated equally. Next, the blur criterion is applied, which is sketched in Fig. \ref{fig:DOF_criteria} as a dashed red line and is defined as the 30\% quantile of the distribution of the calculated metric ($Var(\Delta) \cdot d_\text{B}^2$) over all bubbles. The detected, processed bubbles were linked using trackpy \citep{Allan2021} to calculate the mean diameter $d_\text{m}$ of each bubble and avoid one bubble being counted multiple times when measuring the size distribution of the detached bubbles. Additionally, the standard deviation of $d_\text{m}$ was calculated during the crossing of the ROI and only the bubbles with $Var(d_\text{m}) < 1$ were taken into account for the bubble size distributions. This procedure ensures that misdetected bubbles are excluded from further consideration as far as possible.

\begin{figure}[h]
     \centering
     \begin{minipage}{0.48\textwidth}
        \begin{subfigure}[b]{\textwidth}
            \centering
            \includegraphics[width=\textwidth, frame]{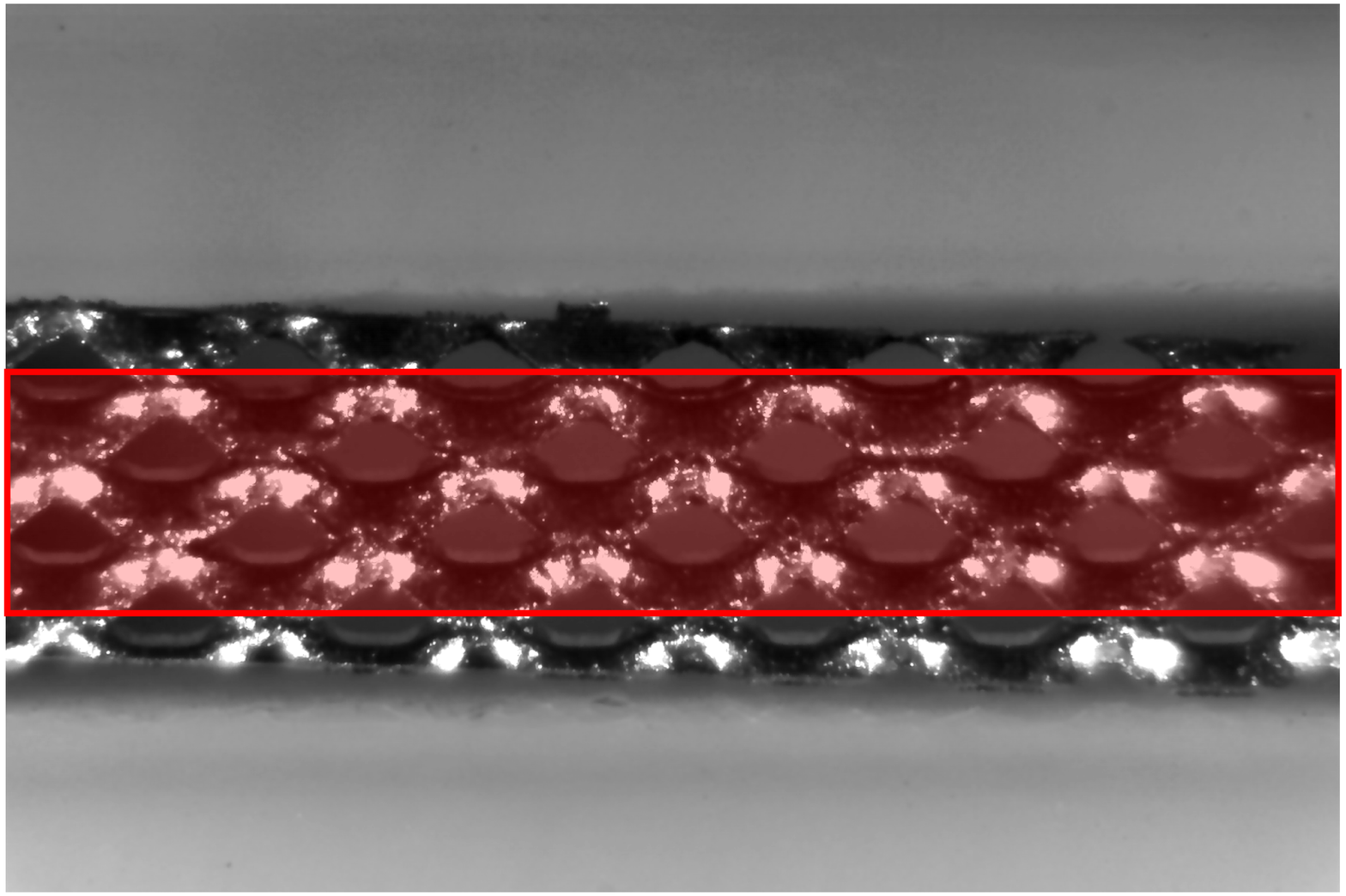}
            \caption{Raw image of the clean electrode}
        \end{subfigure}
    \end{minipage}
    \hfill
    \begin{minipage}{0.48\textwidth}
        \begin{subfigure}[b]{\textwidth}
            \centering
            \includegraphics[width=\textwidth, frame]{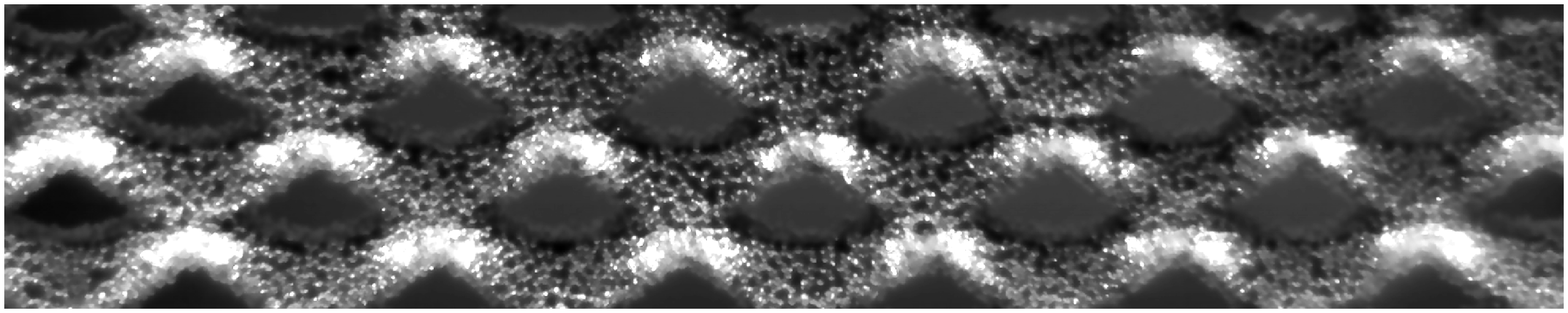}
            \caption{Cropped image of ROI of a fully covered electrode}
        \end{subfigure}
        \begin{subfigure}[b]{\textwidth}
            \centering
            \includegraphics[width=\textwidth, frame]{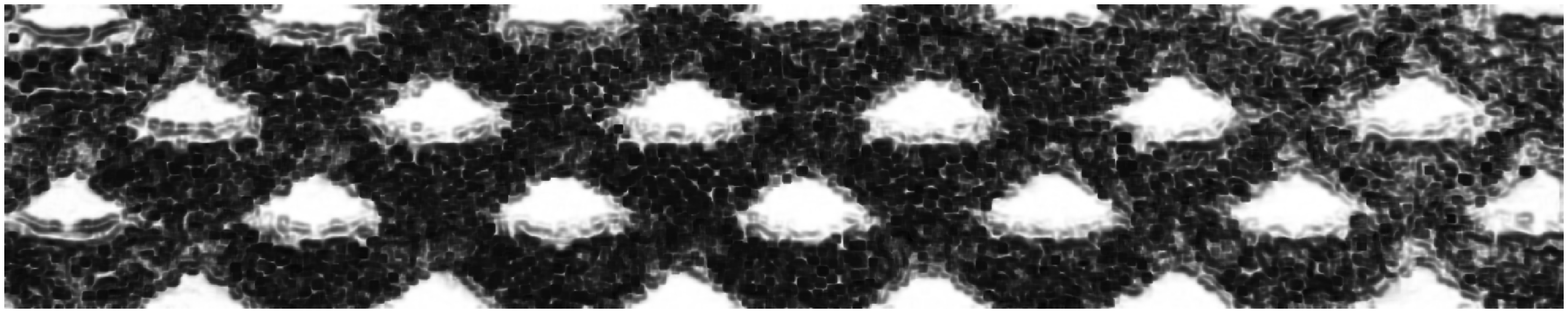}
            \caption{SSIM image of a fully covered electrode}
        \end{subfigure}
        \begin{subfigure}[b]{\textwidth}
            \centering
            \includegraphics[width=\textwidth, frame]{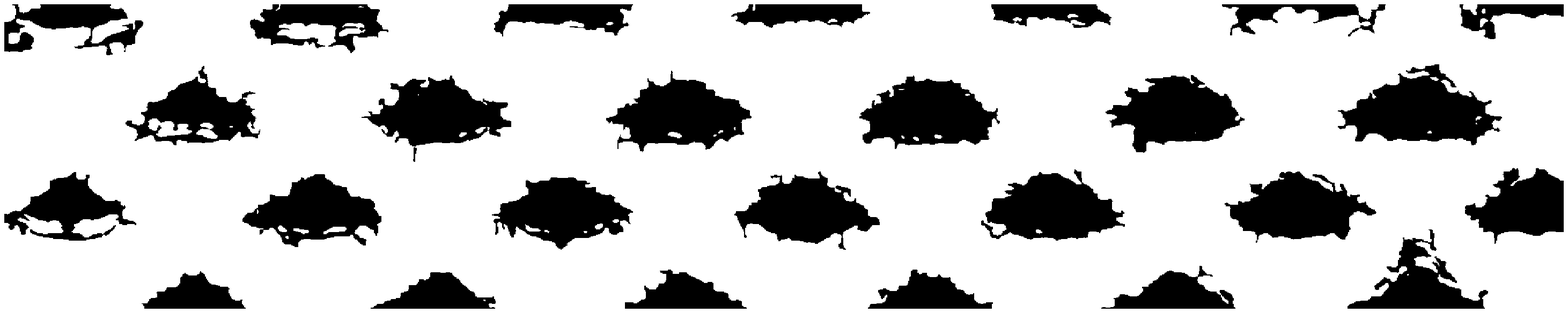}
            \caption{Image with applied threshold to calculate the electrode coverage}
            \label{fig:topview_threshold}
        \end{subfigure}
    \end{minipage}
     \caption{Image processing of top view images consisting of cropping the ROI, calculating the SSIM image of each image by comparing it with the image of the clean electrode and counting the non-zero pixels}
     \label{fig:image_processing_topview}
\end{figure}

The overall electrode coverage was calculated from the top view images using the structural similarity (SSIM) function included in scikit-image \citep{Wang2004, Wang2009}. The SSIM index is a measure of the structural similarity of two images, comparing their luminance, contrast and structure \citep{Wang2004}.
One advantage of the SSIM index is that the result is not influenced by the non-homogeneous illumination of the top view images \citep{Wang2009}. As the laser-cut edges of the electrodes are not representative of the electrode performance, only the central part of the electrode was processed. First, the image of the clean electrode is divided by an average background image. All subsequent frames are then compared to this image using the SSIM function. A sliding window with a size of \SI{9}{pixels}, thus $\approx $ 18 $\times$ \SI{18}{\micro\metre\squared},  is used and the local SSIM index is calculated by comparing it with the same section of the average background image. The size of the sliding window was chosen to guarantee a robust result and to simultaneously resolve the structure of the electrodes. By applying a simple threshold on the calculated SSIM image, the non-zero pixels can be counted; they correspond to those parts of the electrode covered by bubbles. The total coverage  of the electrode $A_\text{cov}$ is calculated using the given values for the electrode porosity $\varepsilon$ (see. Table \ref{tab:electrodes}) and is defined as
\begin{equation}
    A_\text{cov} = \frac{N}{A_\text{im} \cdot (1 - \varepsilon)},
\end{equation}
where $N$ is the total number of non-zero pixels in the threshold image (see Fig. \ref{fig:topview_threshold}) and $A_\text{im}$ is the size of the cropped image. The processing of top view images is sketched in Fig. \ref{fig:image_processing_topview}.

All processed data including example videos are available at \href{https://doi.org/10.14278/rodare.1845}{10.14278/rodare.1845}.

\section{Results and discussion}
\label{sec:results}
\subsection{Electrochemical characterization and performance of electrodes}
\begin{figure}[h]
     \centering
     \begin{subfigure}[t]{0.5\textwidth}
         \centering
         \includegraphics[width=\textwidth]{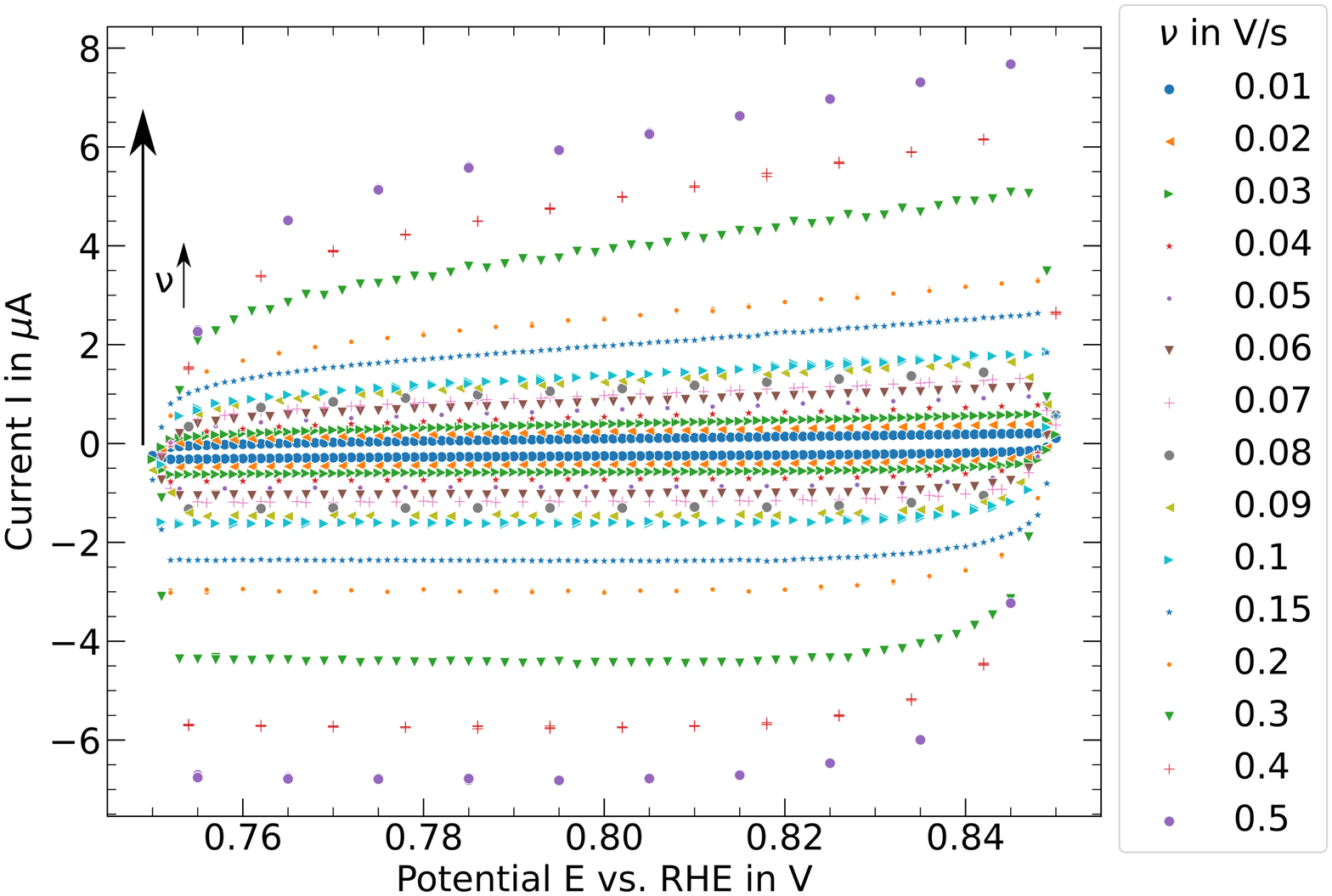}
         \caption{CVs at 15 different scan rates for the electrode \textit{EM\_475}}
         \label{fig:cv_em475}
     \end{subfigure}
     \hfill
     \begin{subfigure}[t]{0.42\textwidth}
         \centering
         \includegraphics[width=\textwidth]{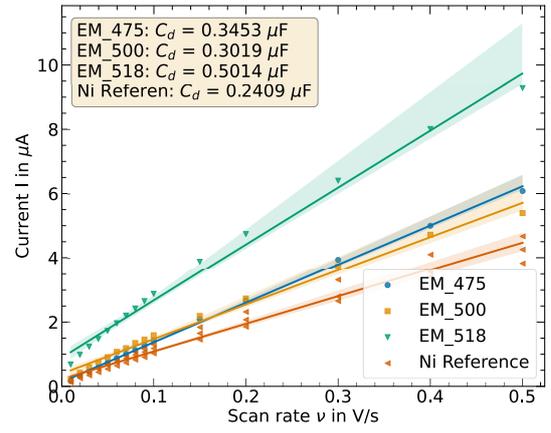}
         \caption{Linear fits of current $I = I_\text{a} + |I_\text{c}|$ over scan rate $\nu$ to calculate double-layer capacitance $C_\text{D}$ of each electrode}
         \label{fig:dlcapacity}
     \end{subfigure}
     \caption{Capacitance measurements to calculate $ECSA$ for all three expanded metal nickel sections using CV}
     \label{fig:electrochemical_characterization}
\end{figure}
The double-layer capacitance $C_\text{D}$ was measured by applying electric potentials at which no significant Faradaic reaction occurs, such that the small current $I$ measured only stems from charging/discharging of the electrode according to the double-layer capacitance $C_\text{D}$ (see Fig.~\ref{fig:dlcapacity}) \citep{Baumann2020}. By running a CV in a wide potential window from \SI{0.25}{\volt} to \SI{0.85}{\volt}, the potential range  of \SI{0.75}{\volt} to \SI{0.85}{\volt}
was obtained.
As can be seen in Fig.~\ref{fig:cv_em475},
nearly rectangular cyclic voltammograms were measured, suggesting that the chosen potential range  
is suitable, and the measured current corresponds 
to charging and discharging of the electrode. 
Thus, the $ECSA$ calculated in accordance with Eq. \ref{eq:ecsa} corresponds to the electrochemically active surface of the electrode. It is important to emphasize that differences in the measured $ECSA$ can be caused by the material properties (e.g. impurities or enhanced roughness) or different geometrical sizes of the uncovered electrode surface. However, all expanded metals show a bigger $ECSA$ compared to the plain nickel foils. 

\begin{figure}[h]
     \centering
     \includegraphics[width=0.5\textwidth]{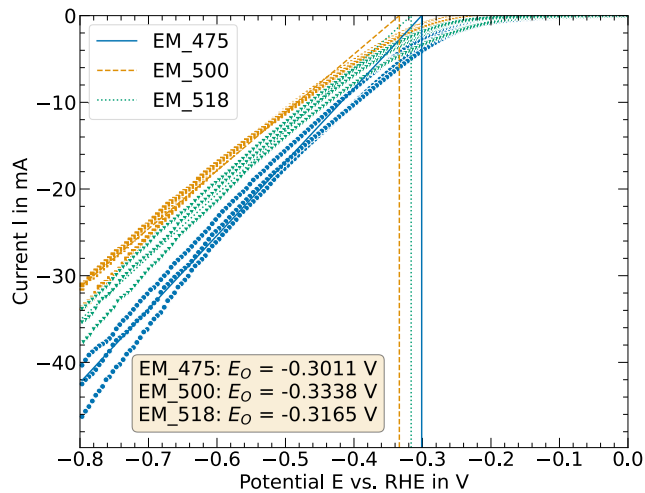}
     \caption{Measured current $I$ during LSV from \SI{0}{\volt} to \SI{-0.8}{\volt} to determine the onset potential $E_\text{O}$ as the intersection of the fitted tangent with \textcolor{black}{the line of $I = \SI{0}{\milli\ampere}$} for three different expanded nickel metals}
     \label{fig:lsv}
\end{figure}

The onset potential for HER, $E_\text{O}$, was determined using LSV (see section \ref{sec:electrochemical_methods}). Fig. \ref{fig:lsv} shows that $E_\text{O}$ is similar for all the expanded metals. Since all the expanded metals were made of nickel with the same purity and were produced with similar manufacturing parameters, this is reasonable. In summary, the main differences between the expanded metals studied lie in the hydraulic diameter $d_\text{h}$ and the $ECSA$. This also takes into account the different thicknesses $t_\text{el}$ of the electrodes (see Table \ref{tab:electrodes}).

One aspect that is crucial for the overall efficiency of the electrolysis is the electrochemical performance of the electrodes, which can be described using the voltage efficiency $\eta_\text{E}$
\begin{align}
    \eta_\text{E} = \frac{|\Delta E^0|}{\Delta E(j)},
\end{align}
where $|\Delta E^0| = \SI{1.23}{\volt}$ is the reversible cell potential for the water splitting reaction under standard conditions and $\Delta E(j)$ is the measured cell potential at a specific current density $j$ \citep{Pang2020}. $\Delta E(j)$ is defined as the sum of $|\Delta E^0|$ and the absolute value of the overpotential losses:
\begin{align}
    \Delta E(j) = |\Delta E^0| + |\eta_\text{HER}| + \eta_\text{OER} + \eta_\Omega + \eta_\text{conc}
    \label{eq:cell potential}
\end{align}
According to \citeauthor{Pang2020} \citep{Pang2020}, the concentration overpotential $\eta_\text{conc}$ is negligible in strong bases. Since all tests were carried out under the same conditions with the same CE, not only  $|\Delta E^0|$ but also the overpotential of the oxygen evolution reaction (OER) $\eta_\text{OER}$ can be assumed to be constant. Thus, a larger measured $\Delta E(j)$ corresponds to greater ohmic overpotential losses $\eta_\Omega$ and HER related losses $\eta_\text{HER}$.

\begin{figure}[h]
     \centering
     \includegraphics[width=0.7\textwidth]{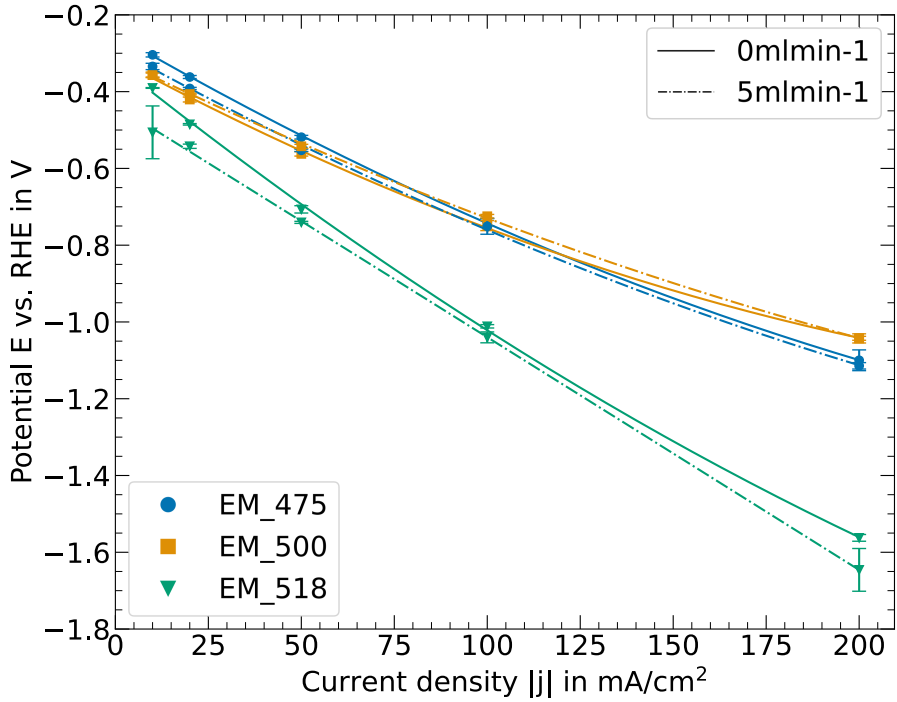}
     \caption{Electrochemical performance of the electrodes \textit{EM\_475}, \textit{EM\_500} and \textit{EM\_518} over a current density range from $|j| = 20$ to \SI{200}{\milli\ampere\per\centi\metre\squared} and at two different flow rates using the averaged potential $\Bar{E}$ vs. RHE of the quasi-steady state region ($t = 8 \dots \SI{10}{\second}$) under galvanostatic conditions}
     \label{fig:EC_performance}
\end{figure}
In Fig. \ref{fig:EC_performance}, the average potential $\Bar{E}$ and standard deviation $\sigma$ of the quasi-steady state region in the time span of $t = 8\dots\SI{10}{\second}$ are plotted over all the current densities $|j|$ studied. In general, greater overpotential losses of the electrode \textit{EM\_518} are observed.  Additionally, for each electrode, two curves are plotted in Fig. \ref{fig:EC_performance} for the cases with and without flow-through. 
It can be stated that the flow-through  has a negligible effect on the potential $E$ for all electrodes and all current densities $j$.
Although this is not what flow-through is intended to achieve in membraneless electrolyzers, 
the phenomenon might be strongly influenced by the specific design of our test cell where the focus is solely on the cathode side,
whereas the flow-by situation at the anodes is far from being optimal. For all the parameters studied, the electrode \textit{EM\_518} shows the highest potential $E(j)$. Since the applied current $I$ was defined by the calculated area $ECSA$, it can be concluded that the overpotential $\eta_\text{HER}$  of the electrode \textit{EM\_518} is greater compared to the other electrodes, assuming constant $\eta_\text{OER}$ at the Pt-CE.

\subsection{Bubble size distributions}
The analysis of the size of detached bubbles $d_\text{32}$ is crucial when designing and
selecting operating conditions for membraneless electrolyzers such
that the overall efficiency and the product purity are
maximized \citep{Davis2019}. When considering the gas crossover within the electrode gap, the detached bubble diameter $d_\text{B}$ has to be minimized in order to reduce the electrode gap. Since a large electrode gap acts as a bottleneck in membraneless electrolyzers compared to other well-established technologies, it must be reduced, though only as long as product purity remains guaranteed.
\begin{figure}[h]
     \centering
     \begin{subfigure}[b]{0.48\textwidth}
         \centering
         \includegraphics[width=\textwidth]{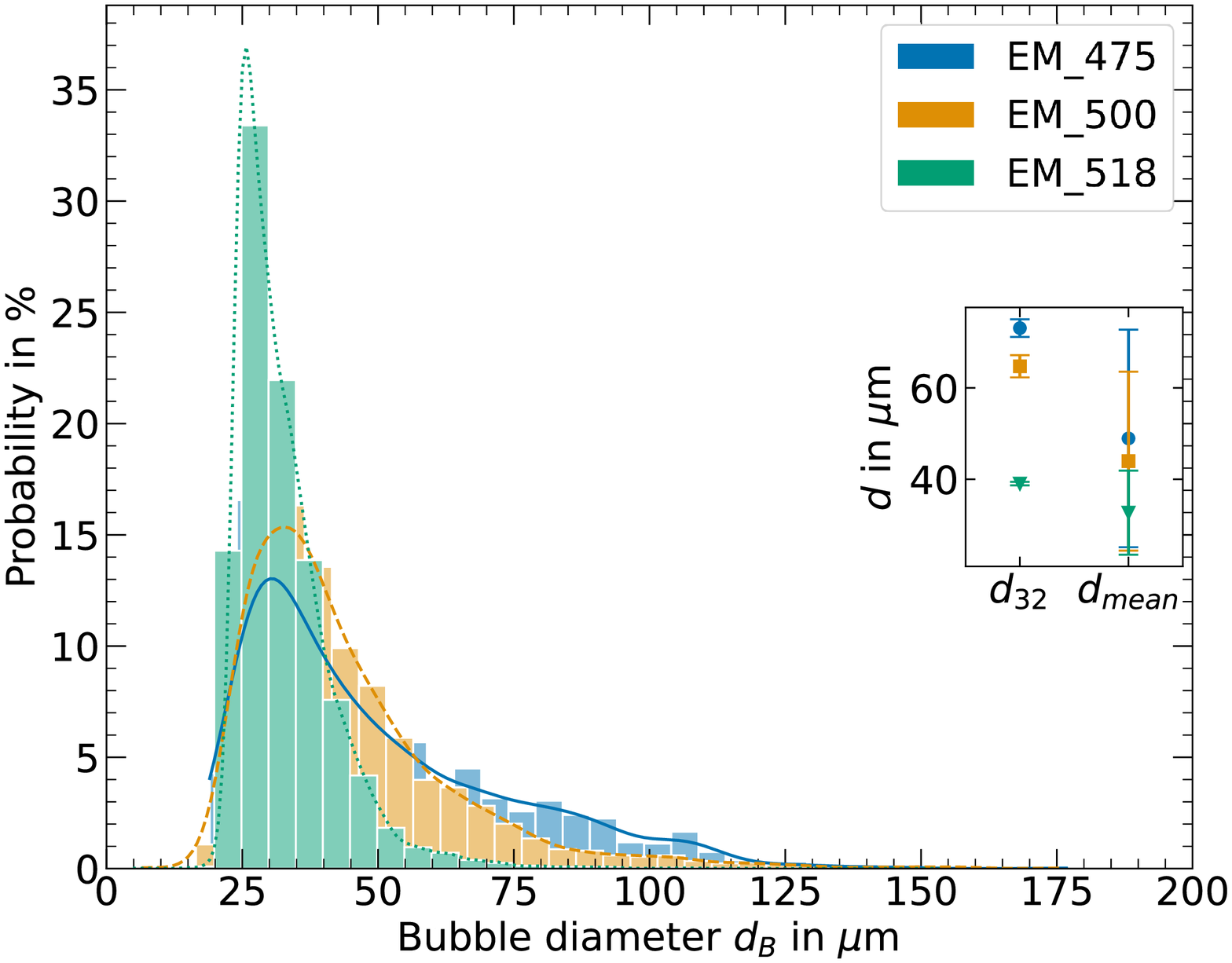}
         \caption{Current density $|j| = \SI{20}{\milli\ampere\per\centi\metre\squared}$ }
         \label{fig:bubblesize_20mAcm^-2}
     \end{subfigure}
     \hfill
     \begin{subfigure}[b]{0.48\textwidth}
         \centering
         \includegraphics[width=\textwidth]{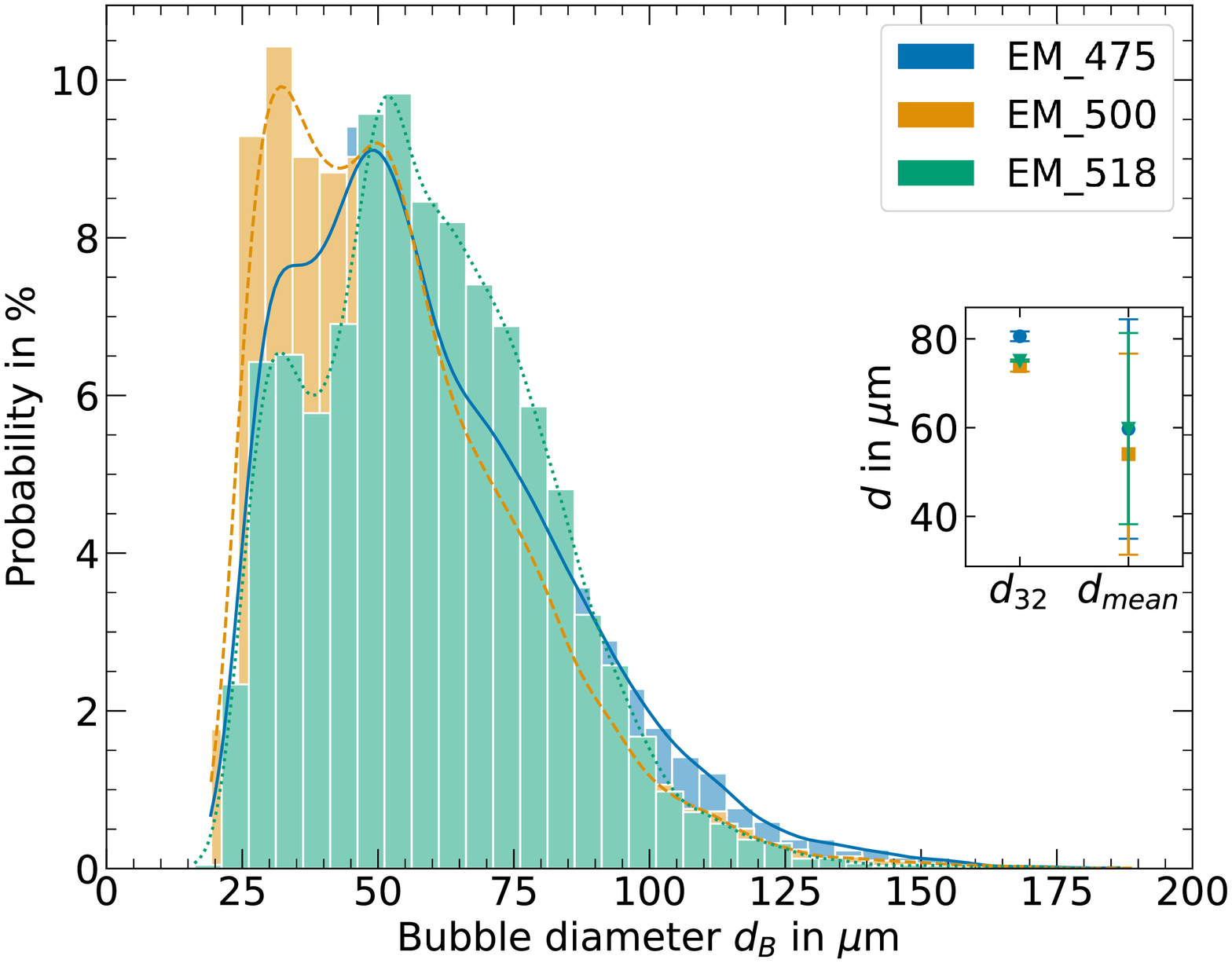}
         \caption{Current density $|j| = \SI{100}{\milli\ampere\per\centi\metre\squared}$ }
         \label{fig:bubblesize_100mAcm^-2}
     \end{subfigure}
     \caption{Size distribution with fitted curves of the detached H$_2$ bubbles at two different current densities $|j|$ and a flow rate of \SI{5}{\milli\litre\per\minute}  for 3 different expanded nickel metals and additional plots of the mean bubble diameter $d_\text{m}$ and Sauter diameter $d_{32}$}
     \label{fig:bubble_distr_dep_electrode}
\end{figure}

In Fig. \ref{fig:bubble_distr_dep_electrode}, the dependence of the WE on the size distribution of detached bubbles is shown when two different current densities are applied, $|j| = \SI{20}{\milli\ampere\per\centi\metre\squared}$ and $|j| = \SI{100}{\milli\ampere\per\centi\metre\squared}$. To provide additional information, the average diameter $d_\text{m}$ and Sauter diameter $d_{32}$ are calculated and plotted for each distribution. Due to the limited spatial resolution, all bubble size distributions are cut off at a bubble diameter of $d_\text{cutoff} \approx \SI{20}{\micro\metre}$ and must be taken into account for all further analysis. Especially at low current densities of $|j| \le \SI{20}{\milli\ampere\per\centi\metre\squared}$ (see  Fig. \ref{fig:bubblesize_20mAcm^-2}), slight differences between the electrodes can be measured. E.g., \textit{EM\_518} shows a smaller mean diameter $d_{m}$ and Sauter diameter $d_{32}$ as well as a narrower bubble size distribution. For a better visualization of the results for all parameter variations, in the following only the Sauter diameter $d_{32}$ will be taken into account.

\begin{figure}[h]
    \centering
    \includegraphics[width=0.6\textwidth]{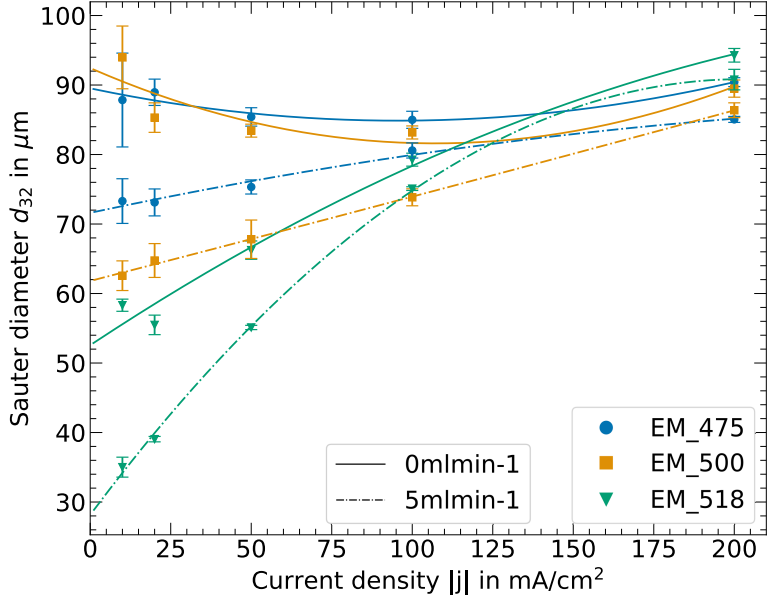}
    \caption{Sauter diameter $d_{32}$ of the detached H$_2$ bubbles as a function of the applied current density $|j|$ and the flow rate $\Dot{V}$}
    \label{fig:bubblesize_sauter}
\end{figure}

For the electrode \textit{EM\_518}, a correlation can be identified between the size distribution of detached bubbles and the current density $|j|$. Increasing current densities $|j|$ lead to increasing \textcolor{black}{Sauter diameters $d_{32}$} (see Fig. \ref{fig:bubblesize_sauter}) and a wider distribution, as shown in Fig. \ref{fig:bubble_distr_dep_electrode}. \textcolor{black}{Increasing bubble sizes were also reported by \citeauthor{Luo2019} \citep{Luo2019} in their study of inexpensive
and efficient electrocatalysts for hydrogen evolution. This phenomenon} can be seen \textcolor{black}{for all studied electrodes} especially clearly if an electrolyte flow is applied (see Fig. \ref{fig:bubblesize_sauter}), even at the low flow rates $\Dot{V}$ applied during the study. The bubble size is drastically reduced \textcolor{black}{(by as much as $\approx$ 40\% at a current density of $|j| = \SI{10}{\milli\ampere\per\centi\metre\squared}$), when the electrolyte flow is applied through the electrodes compared to the no-flow condition. By increasing the electrolyte flow rate $\Dot{V}$, the shear rate at the electrode-bubble interface increases proportionally.} Thus, the bubbles detach faster and at a smaller Sauter diameter $d_{32}$. A similar effect of premature departure due to the applied electrolyte flow was observed by \citeauthor{Zhang2012} \citep{Zhang2012}. Thus, the shear rate is a fundamental parameter to adjust the bubble size and also the product purity in applications such as membraneless electrolyzers. However, the effect of the flow rate falls as the current density $|j|$ increases.
This may be due to the fact that the bubble growth rate and the number of nucleation centers increase with the current density $|j|$. Therefore, at high current densities $|j|$, the low flow rate $\Dot{V}$ applied and thus also the low shear rate no longer influence the bubble growth and detachment so strongly. Here, further experiments \textcolor{black}{including also higher flow rates $\Dot{V}$ are necessary to derive reliable trends at higher current densities $j$.}

\textcolor{black}{However, the electrodes \textit{EM\_518} and \textit{EM\_500} show nearly constant Sauter diameters $d_{32}$ for the no-flow condition compared to the electrode \textit{EM\_518}. In general, the Sauter diameter $d_{32}$ is much less affected by an increase of the current density $j$ without flow. This might indicate that the bubble size is determined by the electrode geometry and electrochemical properties. Here, the electrodes \textit{EM\_518} and \textit{EM\_500} show similar electrochemical performance and the electrode \textit{EM\_518} shows clear differences (see Fig. \ref{fig:EC_performance}).}

It is important to mention that even though the approach based on machine learning leads to better results at higher current densities, many bubbles cannot be detected. This is not due to the algorithm, but to the non-transparent bubble plume detaching at high current densities $|j|$. Within the plume, only big bubbles at the edge of the bubble plumes can usually be segmented. Hence, the bubble size distributions at high current densities $|j|$ \textcolor{black}{are likely to lack} small bubbles. However, the correlation of increasing bubble sizes $d_\text{32}$ with increasing current density $|j|$ \textcolor{black}{remains} valid. Additionally, as the current density $|j|$ increases, not only the size of detached bubbles $d_\text{32}$ but also the bubble layer thickness increases. All of this together leads finally to an increase in the overpotential $\eta_\text{HER}$ \citep{Iwata2021}.
\subsection{Electrode coverage}
\begin{figure}[htb]
     \centering
     \begin{subfigure}[b]{0.48\textwidth}
         \centering
         \includegraphics[width=\textwidth]{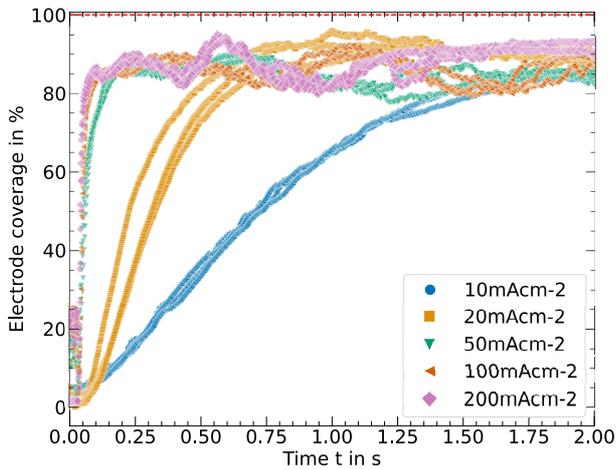}
         \caption{Flow rate $\Dot{V} = \SI{0}{\milli\litre\per\minute}$ }
         \label{fig:coverage_0mlmin^-1}
     \end{subfigure}
     \hfill
     \begin{subfigure}[b]{0.48\textwidth}
         \centering
         \includegraphics[width=\textwidth]{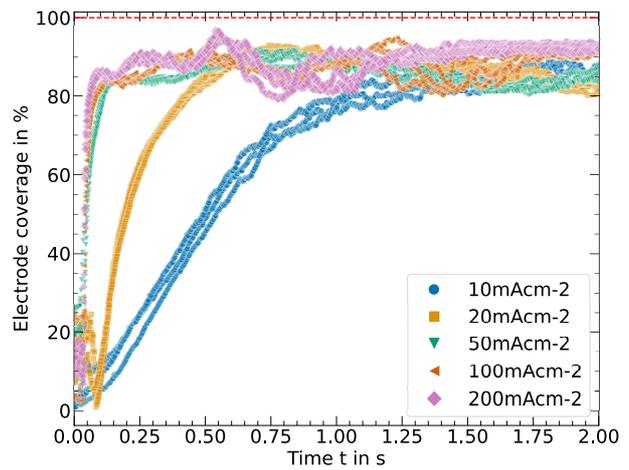}
         \caption{Flow rate $\Dot{V} = \SI{5}{\milli\litre\per\minute}$ }
         \label{fig:coverage_5mlmin^-1}
     \end{subfigure}
     \caption{Dependance of the applied current density $|j|$ and the flow rate $\Dot{V}$ on the coverage of the WE $A_\text{cov}$ over time shown by way of example for the WE \textit{EM\_500}}
     \label{fig:topview_coverage}
\end{figure}

\begin{figure}[htb]
    \centering
    \begin{subfigure}[b]{0.48\textwidth}
         \centering
         \includegraphics[width=\textwidth]{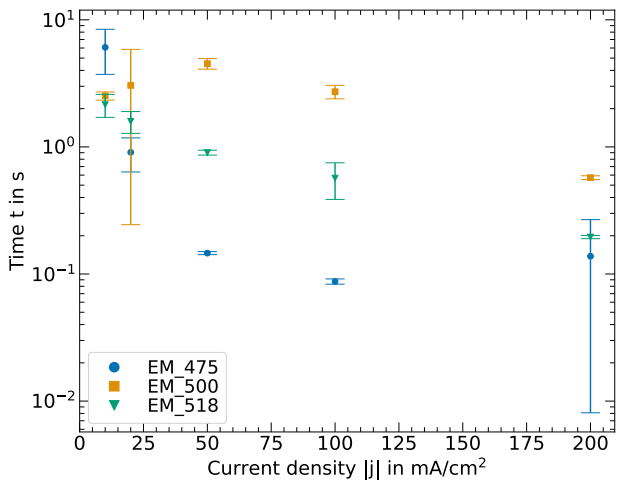}
         \caption{Flow rate $\Dot{V} = \SI{0}{\milli\litre\per\minute}$ }
         \label{fig:full_coverage_0mlmin^-1}
     \end{subfigure}
     \hfill
     \begin{subfigure}[b]{0.48\textwidth}
         \centering
         \includegraphics[width=\textwidth]{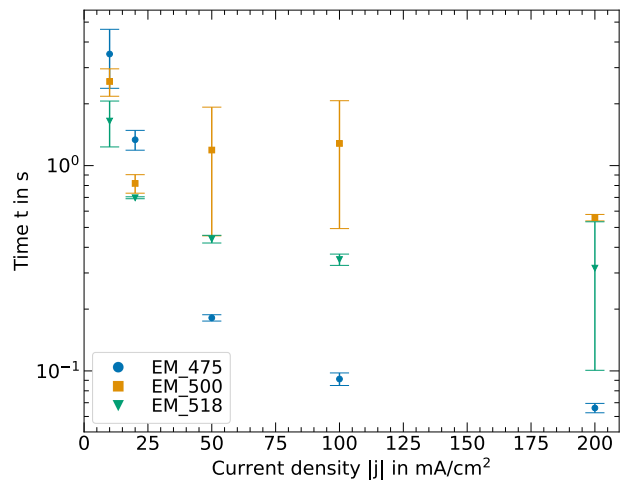}
         \caption{Flow rate $\Dot{V} = \SI{5}{\milli\litre\per\minute}$ }
         \label{fig:full_coverage_5mlmin^-1}
     \end{subfigure}
    \caption{Time $t$ at which the WE surface reached maximal coverage for the different WEs and at electrolyte flowrates of 0 or \SI{5}{\milli\litre\per\minute}}
    \label{fig:electrode_full_cov}
\end{figure}

Fig. \ref{fig:topview_coverage} shows the development of the surface coverage $A_\text{cov}$ over time at different current densities $|j|$ and flow rates $\Dot{V}$ for the WE \textit{EM\_500}. For current densities larger than $|j| \geq \SI{50}{\milli\ampere\per\centi\metre\squared}$ the electrode is nearly fully covered in \textcolor{black}{less than} \SI{0.1}{\second}, whereas it takes significantly longer at low current density $|j|$ (see Fig. \ref{fig:topview_coverage}). This can be observed for both flow rates $\Dot{V}$.  
By plotting the time $t$ at which the maximum coverage of each WE is reached over all the current densities $|j|$ studied (see Fig. \ref{fig:electrode_full_cov}), the fast covering of the electrode at current densities larger than $|j| \geq \SI{50}{\milli\ampere\per\centi\metre\squared}$ can be found for all the electrodes studied. Thus, with an increasing current density $|j|$ the electrode is covered in a shorter timescale. By contrast, the flow rate $\Dot{V}$ shows a negligible effect on the coverage $A_\text{cov}$. However, it is interesting to note that the tendency is for the electrode to be covered more rapidly in the presence of an electrolyte flow. 

The images of the top view have a low spatial resolution, an issue that could not be improved due to difficulties with illumination. Thus, coverages below $A_\text{cov} \approx \SI{20}{\%}$ could not be resolved. Nevertheless, the results show that the current density $|j|$ significantly influences the surface coverage of all the WEs studied. This agrees with Faraday's law, which describes the proportionality between the electric charge and evolved gas, and with the studies by \citeauthor{Vogt2017} \citep{Vogt2017} for low current densities $|j|$. Since the camera and the electrochemical signal were not synchronized, the deviation of the calculated time $t$ in Fig. \ref{fig:electrode_full_cov} reaches values of $\approx$ 50 \%.
In future experiments, this will be avoided by using a trigger signal. Furthermore, since the bubbles rise against the observation window, especially when an electrolyte flow $\Dot{V}$ is applied, the tendency for the coverage time $A_\text{cov}$ to decrease is affected by an additional source of uncertainty: bubbles that are already detached and rising.
\section{Conclusion and Outlook}
This study has introduced a new cell type to perform parametrical studies on various electrodes. Importantly, the cell features a combined analysis of two different perspectives, the top view and side view, that can be used to validate improvements in the bubble dynamics of newly developed electrodes, such as laser-structured or coated electrodes. Moreover, an approach based on machine learning was used to study the size distribution of detached bubbles as a characteristic value for the WE. Additionally, the bubble coverage of the electrode $A_\text{cov}$ was taken into account. Thus, a complete characterization of various electrodes is possible and can help in individual cases to decide which electrode to use. \textcolor{black}{This also makes the cell interesting for the development of new electrodes, since the WE can be characterized holistically.}

The electrodes studied showed similar behaviors for all experimental parameters. Only the expanded \textcolor{black}{nickel} metal \textit{EM\_518} showed higher overpotential $\eta_\text{HER}$. The greatest impact of the applied flow rate $\Dot{V}$ was observed at low current densities ($|j| \le \SI{50}{\milli\ampere\per\centi\metre\squared}$), where the size of detached bubbles $d_\text{32}$ was reduced by as much as $\approx$ 40\%. In further experiments, the flow rate $\Dot{V}$ must be increased to gain a better understanding of the shear rate's influence on the bubble growth and detachment, respectively. \textcolor{black}{Additionally, the current density $j$ is affecting the detached bubble size as well as the electrode coverage $A_\text{cov}$. In general, it can be stated that increasing current densities lead to a faster covering of the electrodes. By increasing the applied current density $j$, the Sauter diameter $d_{32}$ also tends to increase.
However, due to the evolving bubble plumes at high current densities $j$ the results lack in small bubbles entrapped inside these plumes.}

\textcolor{black}{After this proof-of-concept,} the cell will be used to run parametric studies on structured electrodes and on different manufacturing methods. By further optimizing the stardist models and increasing the spatial resolution, it might be possible to investigate bubble dynamics at even higher current densities. 

\section*{Conflicts of interest}
There are no conflicts to declare.

\section*{Acknowledgments}
\textcolor{black}{The Helmholtz Association of German Research Centres (HGF), and German Federal Ministery of Education and Research (BMBF) is gratefully acknowledged for supporting the development of solar powered H$_2$ generation technologies within the frame of the Innovation Pool Project “Clean and Compressed Solar H$_2$“. Furthermore, financial support by the BMBF under contract No. 03SF0672 is gratefully acknowledged. The responsibility for the paper’s content is with the authors.}

\nomenclature[A$d$]{\(d_\text{B}\)}{Bubble diameter in \si{\metre}}
\nomenclature[A$d$]{\(d_{32}\)}{Sauter diameter in \si{\metre}}
\nomenclature[A$d$]{\(d_\text{h}\)}{Hydraulic diameter in \si{\metre}}
\nomenclature[A$d$]{\(d_\text{cutoff}\)}{Cutoff diameter of spatial resolution in \si{\metre}}
\nomenclature[A$N$]{\(N\)}{Number of pixels}
\nomenclature[A$d$]{\(d_\text{m}\)}{Mean diameter in \si{\metre}}
\nomenclature[A$d$]{\(d_\text{in}\)}{Inner diameter in \si{\metre}}
\nomenclature[A$w$]{\(w_\text{mesh}\)}{Mesh width in \si{\metre}}
\nomenclature[A$l$]{\(l_\text{mesh}\)}{Mesh length in \si{\metre}}
\nomenclature[A$w$]{\(w_\text{pore}\)}{Pore width in \si{\metre}}
\nomenclature[A$l$]{\(l_\text{pore}\)}{Pore length in \si{\metre}}
\nomenclature[A$t$]{\(t_\text{el}\)}{Electrode thickness in \si{\metre}}
\nomenclature[A$c$]{\(C_\text{D}\)}{Electrode double-layer capacitance in \si{\farad}}
\nomenclature[A$d$]{\(ECSA\)}{Electrochemically active surface area in \si{\metre\squared}}
\nomenclature[A$E$]{\(E_\text{O}\)}{Onset potential in \si{\volt}}
\nomenclature[A$I$]{\(I\)}{Electrical current in \si{\ampere}}
\nomenclature[A$I$]{\(I_\text{a}\)}{Anodic current in \si{\ampere}}
\nomenclature[A$I$]{\(I_\text{c}\)}{Cathodic current in \si{\ampere}}
\nomenclature[A$c$]{\(c_\text{KOH}\)}{Electrolyte concentration in M}
\nomenclature[A$A$]{\(A_\text{CE}\)}{Counter electrode area in \si{\metre\squared}}
\nomenclature[A$A$]{\(A_\text{cov}\)}{Covered electrode area in \si{\metre\squared}}
\nomenclature[A$A$]{\(A_\text{im}\)}{Image size in px$^2$}
\nomenclature[A$A$]{\(A_\text{pore}\)}{Area of the pore in \si{\metre\squared}}
\nomenclature[A$P$]{\(P_\text{pore}\)}{Perimeter of the pore in \si{\metre}}
\nomenclature[A$A$]{\(A\)}{Electrode area in \si{\metre\squared}}
\nomenclature[A$T$]{\(T\)}{Temperature in \si{\kelvin}}
\nomenclature[A$p$]{\(p\)}{Pressure in \si{\pascal}}
\nomenclature[A$j$]{\(j\)}{Current density in \si{\ampere\per\centi\metre\squared}}
\nomenclature[A$V$]{\(\Dot{V}\)}{Flow rate in \si{\metre\cubed\per\second}}
\nomenclature[A$r$]{\(r_\text{i,j}\)}{Euclidean distance in \si{\metre} (stardist)}
\nomenclature[A$k$]{\(k\)}{Number of radial rays (stardist)}
\nomenclature[A$d$]{\(d_\text{i,j}\)}{Object probability (stardist)}
\nomenclature[A$E$]{\(E\)}{Potential in \si{\volt}}
\nomenclature[A$t$]{\(t\)}{Time in \si{\second}}
\nomenclature[A$E$]{$\Delta E^0$}{Reversible cell potential for the water splitting reaction under standard conditions in \si{\volt}}
\nomenclature[A$E$]{$\Delta E^0$}{Reversible cell potential for the water splitting reaction under standard conditions in \si{\volt}}

\nomenclature[b]{\(\varepsilon\)}{Porosity in \%}
\nomenclature[b]{\(\eta_\text{E}\)}{Voltage efficiency in \%}
\nomenclature[b]{\(\eta_\text{HER}\)}{Overpotential of the hydrogen evolution reaction in \si{\volt}}
\nomenclature[b]{\(\eta_\text{OER}\)}{Overpotential of the oxygen evolution reaction in \si{\volt}}
\nomenclature[b]{\(\eta_\Omega\)}{Ohmic overpotential in \si{\volt}}
\nomenclature[b]{\(\eta_\text{conc}\)}{Concentration overpotential in \si{\volt}}
\nomenclature[b]{\(\sigma\)}{Standard deviation}
\nomenclature[b]{\(\nu\)}{Scan rate in \si{\volt\per\second}}
\nomenclature[]{\(\)}{}
\printnomenclature[2cm]

\bibliographystyle{unsrtnat}
\bibliography{references}

\begin{thebibliography}{54}
\providecommand{\natexlab}[1]{#1}
\providecommand{\url}[1]{\texttt{#1}}
\expandafter\ifx\csname urlstyle\endcsname\relax
  \providecommand{\doi}[1]{doi: #1}\else
  \providecommand{\doi}{doi: \begingroup \urlstyle{rm}\Url}\fi

\bibitem[van Renssen(2020)]{Renssen2020}
S.~van Renssen.
\newblock The hydrogen solution?
\newblock \emph{Nature Climate Change}, 10\penalty0 (9):\penalty0 799--801,
  2020.
\newblock \doi{10.1038/s41558-020-0891-0}.

\bibitem[van~der Spek et~al.(2022)van~der Spek, Banet, Bauer, Gabrielli,
  Goldthorpe, Mazzotti, Munkejord, Røkke, Shah, Sunny, Sutter, Trusler, and
  Gazzani]{Spek2022}
M.~van~der Spek, C.~Banet, C.~Bauer, P.~Gabrielli, W.~Goldthorpe, M.~Mazzotti,
  S.~T. Munkejord, N.~A. Røkke, N.~Shah, N.~Sunny, D.~Sutter, J.~M. Trusler,
  and M.~Gazzani.
\newblock Perspective on the hydrogen economy as a pathway to reach net-zero
  co2 emissions in europe.
\newblock \emph{Energy Environmental Science}, 15:\penalty0 1034--1077, 2022.
\newblock \doi{10.1039/D1EE02118D}.

\bibitem[Staffell et~al.(2019)Staffell, Scamman, Abad, Balcombe, Dodds, Ekins,
  Shah, and Ward]{Staffell2019}
I.~Staffell, D.~Scamman, A.~V. Abad, P.~Balcombe, P.~E. Dodds, P.~Ekins,
  N.~Shah, and K.~R. Ward.
\newblock The role of hydrogen and fuel cells in the global energy system.
\newblock \emph{Energy {\&} Environmental Science}, 12\penalty0 (2):\penalty0
  463--491, 2019.
\newblock \doi{10.1039/c8ee01157e}.

\bibitem[Smolinka and Garche(2021)]{smolinka2021electrochemical}
T.~Smolinka and J.~Garche.
\newblock \emph{Electrochemical Power Sources: Fundamentals, Systems, and
  Applications: Hydrogen Production by Water Electrolysis}.
\newblock Elsevier, 2021.
\newblock ISBN 9780128194256.

\bibitem[Milani et~al.(2020)Milani, Kiani, and McNaughton]{Milani2020}
D.~Milani, A.~Kiani, and R.~McNaughton.
\newblock Renewable-powered hydrogen economy from australia{\textquotesingle}s
  perspective.
\newblock \emph{International Journal of Hydrogen Energy}, 45\penalty0
  (46):\penalty0 24125--24145, 2020.
\newblock \doi{10.1016/j.ijhydene.2020.06.041}.

\bibitem[Sazali(2020)]{Sazali2020}
N.~Sazali.
\newblock Emerging technologies by hydrogen: A review.
\newblock \emph{International Journal of Hydrogen Energy}, 45\penalty0
  (38):\penalty0 18753--18771, 2020.
\newblock \doi{10.1016/j.ijhydene.2020.05.021}.

\bibitem[{European Comission}(2020)]{EUstrategy}
{European Comission}.
\newblock A hydrogen strategy for a climate-neutral europe.
\newblock
  \url{https://ec.europa.eu/energy/sites/ener/files/hydrogen_strategy.pdf},
  2020.
\newblock Accessed: 2022-06-14.

\bibitem[Angulo et~al.(2020)Angulo, van~der Linde, Gardeniers, Modestino, and
  Rivas]{Angulo2020}
A.~Angulo, P.~van~der Linde, H.~Gardeniers, M.~Modestino, and D.~F. Rivas.
\newblock Influence of bubbles on the energy conversion efficiency of
  electrochemical reactors.
\newblock \emph{Joule}, 4\penalty0 (3):\penalty0 555--579, 2020.
\newblock \doi{10.1016/j.joule.2020.01.005}.

\bibitem[Zhao et~al.(2019)Zhao, Ren, and Luo]{zhao2019gas}
X.~Zhao, H.~Ren, and L.~Luo.
\newblock Gas bubbles in electrochemical gas evolution reactions.
\newblock \emph{Langmuir}, 35\penalty0 (16):\penalty0 5392--5408, 2019.
\newblock \doi{10.1021/acs.langmuir.9b00119}.

\bibitem[Leistra and Sides(1987)]{leistra1987voltage}
J.~A. Leistra and P.~J. Sides.
\newblock Voltage components at gas evolving electrodes.
\newblock \emph{Journal of the Electrochemical Society}, 134\penalty0
  (10):\penalty0 2442, 1987.
\newblock \doi{10.1149/1.2100218}.

\bibitem[Lake et~al.(2022)Lake, Soto, and Varanasi]{lake2022impact}
J.~R. Lake, \'A.~M. Soto, and K.~K. Varanasi.
\newblock Impact of bubbles on electrochemically active surface area of
  microtextured gas-evolving electrodes.
\newblock \emph{Langmuir}, 2022.
\newblock \doi{10.1021/acs.langmuir.2c00035}.

\bibitem[Esposito(2017)]{Esposito2017}
D.~V. Esposito.
\newblock Membraneless {E}lectrolyzers for {L}ow-{C}ost {H}ydrogen {P}roduction
  in a {R}enewable {E}nergy {F}uture.
\newblock \emph{Joule}, 1:\penalty0 651--658, 2017.
\newblock \doi{10.1016/j.joule.2017.07.003}.

\bibitem[Gillespie et~al.(2015)Gillespie, van~der Merwe, and
  Kriek]{Gillespie2015}
M.~I. Gillespie, F.~van~der Merwe, and R.~J. Kriek.
\newblock Performance evaluation of a membraneless divergent
  electrode-flow-through ({DEFT}) alkaline electrolyser based on optimisation
  of electrolytic flow and elctrode gap.
\newblock \emph{Journal of Power Sources}, 293:\penalty0 228--235, 2015.
\newblock \doi{10.1016/j.jpowsour.2015.05.077}.

\bibitem[Gillespie and Kriek(2017)]{Gillespie2017}
M.~I. Gillespie and R.~J. Kriek.
\newblock Hydrogen production from a rectangular horizontal filter press
  {D}ivergent {E}lectrode-{F}low-{T}hrough ({DEFT}) alkaline electrolysis
  stack.
\newblock \emph{Journal of Power Sources}, 372:\penalty0 252--259, 2017.
\newblock \doi{10.1016/j.jpowsour.2017.10.080}.

\bibitem[Gillespie and Kriek(2018)]{Gillespie2018}
M.~I. Gillespie and R.~J. Kriek.
\newblock Scalable hydrogen production from a mono-circular filter press
  {D}ivergent {E}lectrode-{F}low-{T}hrough alkaline electrolysis stack.
\newblock \emph{Journal of Power Sources}, 397:\penalty0 204--213, 2018.
\newblock \doi{10.1016/j.jpowsour.2018.07.026}.

\bibitem[Bui et~al.(2020)Bui, Davis, and Esposito]{Bui2020}
J.~C. Bui, J.~T. Davis, and D.~V. Esposito.
\newblock 3{D}-{P}rinted electrodes for membraneless water electrolysis.
\newblock \emph{Sustainable Energy \& Fuels}, 4:\penalty0 213--225, 2020.
\newblock \doi{10.1039/c9se00710e}.

\bibitem[Davis et~al.(2019)Davis, Brown, Pang, and Esposito]{Davis2019}
J.~T. Davis, D.~E. Brown, X.~Pang, and D.~V. Esposito.
\newblock High {S}peed {V}ideo {I}nvestigation of {B}ubble {D}ynamics and
  {C}urrent {D}ensity {D}istributions in {M}embraneless {E}lectrolyzers.
\newblock \emph{Journal of the Electrochemical Society}, 166:\penalty0
  F312--F321, 2019.
\newblock \doi{10.1149/2.0961904jes}.

\bibitem[O'Neil et~al.(2016)O'Neil, Christian, Brown, and Esposito]{Neil2016}
G.~D. O'Neil, C.~D. Christian, D.~E. Brown, and D.~V. Esposito.
\newblock Hydrogen {P}roduction with a {S}imple and {S}calable {M}embraneless
  {E}lectrolyzer.
\newblock \emph{Journal of The Electrochemical Society}, 163:\penalty0
  F3012--F3019, 2016.
\newblock \doi{10.1149/2.0021611jes}.

\bibitem[Pang et~al.(2020)Pang, Davis, Ill, and Esposito]{Pang2020}
X.~Pang, J.~T. Davis, A.~D.~Harvey Ill, and D.~V. Esposito.
\newblock Framework for evaluating the performance limits of membraneless
  electrolyzers.
\newblock \emph{Energy \& Environmental Science}, 13:\penalty0 3663--3678,
  2020.
\newblock \doi{10.1039/d0ee02268c}.

\bibitem[Hashemi et~al.(2015)Hashemi, Modestino, and Psaltis]{Hashemi2015}
S.~M.~H. Hashemi, M.~A. Modestino, and D.~Psaltis.
\newblock A membrane-less electrolyzer for hydrogen production across the p{H}
  scale.
\newblock \emph{Energy \& Environmental Science}, 8:\penalty0 2003--2009, 2015.
\newblock \doi{10.1039/c5ee00083a}.

\bibitem[Hashemi et~al.(2019)Hashemi, Karnakov, Hadikhani, Chinello, Litvinov,
  Moser, Koumoutsakos, and Psaltis]{Hashemi2019}
S.~M.~H. Hashemi, P.~Karnakov, P.~Hadikhani, E.~Chinello, S.~Litvinov,
  C.~Moser, P.~Koumoutsakos, and D.~Psaltis.
\newblock A versatile and membrane-less electrochemical reactor for the
  electrolysis of water and brine.
\newblock \emph{Energy \& Environmental Science}, 12\penalty0 (5):\penalty0
  1592--1604, 2019.
\newblock \doi{10.1039/c9ee00219g}.

\bibitem[Dotan et~al.(2019)Dotan, Landman, Sheehan, Malviya, Shter, Grave,
  Arzi, Yehudai, Halabi, Gal, Hadari, Cohen, Rothschild, and Grader]{Dotan2019}
H.~Dotan, A.~Landman, S.~W. Sheehan, K.~D. Malviya, G.~E. Shter, D.~A. Grave,
  Z.~Arzi, N.~Yehudai, M.~Halabi, N.~Gal, N.~Hadari, C.~Cohen, A.~Rothschild,
  and G.~S. Grader.
\newblock Decoupled hydrogen and oxygen evolution by a two-step
  electrochemical{\textendash}chemical cycle for efficient overall water
  splitting.
\newblock \emph{Nature Energy}, 4\penalty0 (9):\penalty0 786--795, 2019.
\newblock \doi{10.1038/s41560-019-0462-7}.

\bibitem[Yan et~al.(2021)Yan, Biemolt, Zhao, Zhao, Cao, Yang, Wu, Rothenberg,
  and Yan]{Yan2021}
X.~Yan, J.~Biemolt, K.~Zhao, Y.~Zhao, X.~Cao, Y.~Yang, X.~Wu, G.~Rothenberg,
  and N.~Yan.
\newblock A membrane-free flow electrolyzer operating at high current density
  using earth-abundant catalysts for water splitting.
\newblock \emph{Nature Communications}, 12\penalty0 (1), 2021.
\newblock \doi{10.1038/s41467-021-24284-5}.

\bibitem[Solovey et~al.(2018{\natexlab{a}})Solovey, Zipunnikov, Shevchenko,
  Vorobjova, and Kotenko]{Solovey2018a}
V.~Solovey, M.~Zipunnikov, A.~Shevchenko, I.~Vorobjova, and A.~Kotenko.
\newblock Energy {E}ffective {M}embrane-less {T}echnology for {H}igh {P}ressure
  {H}ydrogen {E}lectro-chemical {G}eneration.
\newblock \emph{French-Ukrainian Journal of Chemistry}, 6:\penalty0 151--156,
  2018{\natexlab{a}}.
\newblock \doi{10.17721/fujcV6I1P151-156}.

\bibitem[Solovey et~al.(2018{\natexlab{b}})Solovey, Khiem, Zipunnikov, and
  Shevchenko]{Solovey2018b}
V.~Solovey, N.~T. Khiem, M.~Mykolaevich Zipunnikov, and A.~Shevchenko.
\newblock Improvement of the {M}embrane - less {E}lectrolysis {T}echnology for
  {H}ydrogen and {O}xygen {G}eneration.
\newblock \emph{French-Ukrainian Journal of Chemistry}, 6:\penalty0 73--79,
  2018{\natexlab{b}}.
\newblock \doi{10.17721/fujcV6I2P73-79}.

\bibitem[Solovey et~al.(2021)Solovey, Shevchenko, Zipunnikov, Kotenko, Khiem,
  Tri, and Hai]{Solovey2021}
V.~Solovey, A.~Shevchenko, M.~Zipunnikov, A.~L. Kotenko, N.~T. Khiem, B.~D.
  Tri, and T.~T. Hai.
\newblock Development of high pressure membraneless alkaline electrolyzer.
\newblock \emph{International Journal of Hydrogen Energy}, 2021.
\newblock \doi{10.1016/j.ijhydene.2021.01.209}.

\bibitem[Rajaei et~al.(2021)Rajaei, Rajora, and Haverkort]{Rajaei2021}
H.~Rajaei, A.~Rajora, and J.~W. Haverkort.
\newblock Design of membraneless gas-evolving flow-through porous electrodes.
\newblock \emph{Journal of Power Sources}, 491:\penalty0 229364, 2021.
\newblock \doi{10.1016/j.jpowsour.2020.229364}.

\bibitem[X. et~al.(2018)X., Baczyzmalski, Cierpka, Mutschke, and
  Eckert]{Yang2018}
X., D.~Baczyzmalski, C.~Cierpka, G.~Mutschke, and K.~Eckert.
\newblock Marangoni convection at electrogenerated hydrogen bubbles.
\newblock \emph{Physical Chemistry Chemical Physics}, 20\penalty0
  (17):\penalty0 11542--11548, 2018.
\newblock \doi{10.1039/c8cp01050a}.

\bibitem[Bashkatov et~al.(2019)Bashkatov, Hossain, Yang, Mutschke, and
  Eckert]{Bashkatov2019}
A.~Bashkatov, S.~S. Hossain, X.~Yang, G.~Mutschke, and K.~Eckert.
\newblock Oscillating {H}ydrogen {B}ubbles at {P}t {M}icroelectrodes.
\newblock \emph{Physical Review Letters}, 132, 2019.
\newblock \doi{10.1103/PhysRevLett.123.214503}.

\bibitem[Hossain et~al.(2020)Hossain, Mutschke, Bashkatov, and
  Eckert]{Hossain2020}
S.~S. Hossain, G.~Mutschke, A.~Bashkatov, and K.~Eckert.
\newblock The thermocapillary effect on gas bubbles growing on electrodes of
  different sizes.
\newblock \emph{Electrochimica Acta}, 353:\penalty0 136461, 2020.
\newblock \doi{10.1016/j.electacta.2020.136461}.

\bibitem[Meulenbroek et~al.(2021)Meulenbroek, Vreman, and
  Deen]{Meulenbroek2021}
A.~M. Meulenbroek, A.~W. Vreman, and N.~G. Deen.
\newblock Competing marangoni effects form a stagnant cap on the interface of a
  hydrogen bubble attached to a microelectrode.
\newblock \emph{Electrochimica Acta}, page 138298, 2021.
\newblock \doi{10.1016/j.electacta.2021.138298}.

\bibitem[Hossain et~al.(2022)Hossain, Bashkatov, Yang, Mutschke, and
  Eckert]{hossain2022}
S.~S. Hossain, A.~Bashkatov, X.~Yang, G.~Mutschke, and K.~Eckert.
\newblock Force balance of hydrogen bubbles growing and oscillating on a
  microelectrode.
\newblock \emph{Physical Review E}, 106:\penalty0 035105, 2022.
\newblock \doi{10.1103/PhysRevE.106.035105}.

\bibitem[Bashkatov et~al.(2022{\natexlab{a}})Bashkatov, Hossain, Mutschke,
  Yang, Rox, Weidinger, and Eckert]{bashkatov2022}
A.~Bashkatov, S.~S. Hossain, G.~Mutschke, X.~Yang, H.~Rox, I.~M. Weidinger, and
  K.~Eckert.
\newblock On the growth regimes of hydrogen bubbles at microelectrodes.
\newblock \emph{Accepted in Physical Chemistry Chemical Physics},
  2022{\natexlab{a}}.

\bibitem[Zhang et~al.(2010)Zhang, Merrill, and Logan]{Zhang2010}
Y.~Zhang, M.~D. Merrill, and B.~E. Logan.
\newblock The use and optimization of stainless steel mesh cathodes in
  microbial electrolysis cells.
\newblock \emph{International Journal of Hydrogen Energy}, 35\penalty0
  (21):\penalty0 12020--12028, 2010.
\newblock \doi{10.1016/j.ijhydene.2010.08.064}.

\bibitem[Li et~al.(2021)Li, Jiang, Zhang, Feng, Zhang, Yao, and Wang]{Li2021c}
L.~Li, W.~Jiang, G.~Zhang, D.~Feng, C.~Zhang, W.~Yao, and Z.~Wang.
\newblock Efficient mesh interface engineering: Insights from bubble dynamics
  in electrocatalysis.
\newblock \emph{{ACS} Applied Materials {\&} Interfaces}, 2021.
\newblock \doi{10.1021/acsami.1c07637}.

\bibitem[Lee et~al.(2021)Lee, Cho, Kim, Lee, Lee, Lee, Kim, Kim, Yi, and
  Cho]{Lee2021b}
H.~I. Lee, H.-S. Cho, M.~Kim, J.~Hun Lee, C.~Lee, S.~Lee, S.-K. Kim, C.-H. Kim,
  K.~B. Yi, and W.-C. Cho.
\newblock The structural effect of electrode mesh on hydrogen evolution
  reaction performance for alkaline water electrolysis.
\newblock \emph{Frontiers in Chemistry}, 9, 2021.
\newblock \doi{10.3389/fchem.2021.787787}.

\bibitem[Reyssat(2014)]{Reyssat2014}
E.~Reyssat.
\newblock Drops and bubbles in wedges.
\newblock \emph{Journal of Fluid Mechanics}, 748:\penalty0 641--662, 2014.
\newblock \doi{10.1017/jfm.2014.201}.

\bibitem[Yang et~al.(2022)Yang, Li, Yang, Lan, Liu, Fu, Zhang, Liao, and
  Zhu]{Yang2022}
Y.~Yang, J.~Li, Y.~Yang, L.~Lan, R.~Liu, Q.~Fu, L.~Zhang, Q.~Liao, and X.~Zhu.
\newblock Gradient porous electrode-inducing bubble splitting for highly
  efficient hydrogen evolution.
\newblock \emph{Applied Energy}, 307:\penalty0 118278, 2022.
\newblock \doi{10.1016/j.apenergy.2021.118278}.

\bibitem[Vogt(2011)]{Vogt2011}
H.~Vogt.
\newblock On the gas-evolution efficiency of electrodes {I} - {T}heoretical.
\newblock \emph{Electrochimica Acta}, 56\penalty0 (3):\penalty0 1409--1416,
  2011.
\newblock \doi{10.1016/j.electacta.2010.08.101}.

\bibitem[Eigeldinger and Vogt(2000)]{Eigeldinger2000}
J.~Eigeldinger and H.~Vogt.
\newblock The bubble coverage of gas-evolving electrodes in a flowing
  electrolyte.
\newblock \emph{Electrochimica Acta}, 45\penalty0 (27):\penalty0 4449--4456,
  2000.
\newblock ISSN 0013-4686.
\newblock \doi{10.1016/S0013-4686(00)00513-2}.

\bibitem[Vogt and Stephan(2015)]{Vogt2015}
H.~Vogt and K.~Stephan.
\newblock Local microprocesses at gas-evolving electrodes and their influence
  on mass transfer.
\newblock \emph{Electrochimica Acta}, 155:\penalty0 348--356, 2015.
\newblock \doi{10.1016/j.electacta.2015.01.008}.

\bibitem[Vogt(2017)]{Vogt2017}
H.~Vogt.
\newblock The quantities affecting the bubble coverage of gas-evolving
  electrodes.
\newblock \emph{Electrochimica Acta}, 235:\penalty0 495--499, 2017.
\newblock \doi{10.1016/j.electacta.2017.03.116}.

\bibitem[Bashkatov et~al.(2022{\natexlab{b}})Bashkatov, Hossain, Rox, Yang,
  Mutschke, and Eckert]{Bashkatov2022FC3}
A.~Bashkatov, S.~S. Hossain, H.~Rox, X.~Yang, G.~Mutschke, and K.~Eckert.
\newblock Dynamics of single $h_2$ bubbles during water electrolysis.
\newblock \emph{FC³ Fuel Cell Conference. Chemnitz, 31.05./01.06.2022},
  2022{\natexlab{b}}.

\bibitem[Skibińska et~al.(2022)Skibińska, Kornaus, Yang, Kutyła, Wojnicki,
  and Żabiński]{Skibinska2022}
K.~Skibińska, K.~Kornaus, X.~Yang, D.~Kutyła, M.~Wojnicki, and P.~Żabiński.
\newblock One-step synthesis of the hydrophobic conical co-fe structures-the
  comparison of their active areas and electrocatalytic properties.
\newblock \emph{Electrochimica Acta}, 415:\penalty0 140127, 2022.
\newblock \doi{https://doi.org/10.1016/j.electacta.2022.140127}.

\bibitem[Schmidt et~al.(2018)Schmidt, Weigert, Broaddus, and
  Myers]{Schmidt2018}
U.~Schmidt, M.~Weigert, C.~Broaddus, and G.~Myers.
\newblock Cell detection with star-convex polygons.
\newblock In \emph{Medical Image Computing and Computer Assisted Intervention -
  {MICCAI} 2018 - 21st International Conference, Granada, Spain, September
  16-20, 2018, Proceedings, Part {II}}, pages 265--273, 2018.
\newblock \doi{10.1007/978-3-030-00934-2-30}.

\bibitem[Weigert et~al.(2020)Weigert, Schmidt, Haase, Sugawara, and
  Myers]{Weigert2020}
M.~Weigert, U.~Schmidt, R.~Haase, K.~Sugawara, and G.~Myers.
\newblock Star-convex polyhedra for 3d object detection and segmentation in
  microscopy.
\newblock In \emph{The IEEE Winter Conference on Applications of Computer
  Vision (WACV)}, 2020.
\newblock \doi{10.1109/WACV45572.2020.9093435}.

\bibitem[Hessenkemper et~al.(2022)Hessenkemper, Starke, Atassi, Ziegenhein, and
  Lucas]{Hessenkemper2022}
H.~Hessenkemper, S.~Starke, Y.~Atassi, T.~Ziegenhein, and D.~Lucas.
\newblock Bubble identification from images with machine learning methods.
\newblock \emph{International Journal of Multiphase Flow}, 155:\penalty0
  104169, 2022.
\newblock \doi{10.1016/j.ijmultiphaseflow.2022.104169}.

\bibitem[Allan et~al.(2021)Allan, Caswell, Keim, van~der Wel, and
  Verweij]{Allan2021}
D.~B. Allan, T.~Caswell, N.~C. Keim, C.~M. van~der Wel, and R.~W. Verweij.
\newblock soft-matter/trackpy: Trackpy v0.5.0.
\newblock \emph{Zenodo}, 2021.
\newblock \doi{10.5281/zenodo.4682814}.

\bibitem[Wang et~al.(2004)Wang, Bovik, Sheikh, and Simoncelli]{Wang2004}
Z.~Wang, A.~C. Bovik, H.~R. Sheikh, and E.~P. Simoncelli.
\newblock Image quality assessment: from error visibility to structural
  similarity.
\newblock \emph{IEEE transactions on image processing}, 13\penalty0
  (4):\penalty0 600--612, 2004.
\newblock \doi{10.1109/TIP.2003.819861}.

\bibitem[Wang and Bovik(2009)]{Wang2009}
Z.~Wang and A.~C. Bovik.
\newblock Mean squared error: Love it or leave it? a new look at signal
  fidelity measures.
\newblock \emph{IEEE signal processing magazine}, 26\penalty0 (1):\penalty0
  98--117, 2009.
\newblock \doi{10.1109/MSP.2008.930649}.

\bibitem[Baumann et~al.(2020)Baumann, Rauscher, Bernäcker, Zwahr,
  Weißgärber, Röntzsch, and Lasagni]{Baumann2020}
R.~Baumann, T.~Rauscher, C.~I. Bernäcker, C.~Zwahr, T.~Weißgärber,
  L.~Röntzsch, and A.~F. Lasagni.
\newblock Laser structuring of open cell metal foams for micro scale surface
  enlargement.
\newblock \emph{Journal of Laser Micro/Nanoengineering}, 2020.
\newblock \doi{10.2961/jlmn.2020.02.2010}.

\bibitem[Luo et~al.(2019)Luo, Tang, Khan, Yu, Cheng, Zou, and Liu]{Luo2019}
Y.~Luo, L.~Tang, U.~Khan, Q.~Yu, H.-M. Cheng, X.~Zou, and B.~Liu.
\newblock Morphology and surface chemistry engineering toward {pH}-universal
  catalysts for hydrogen evolution at high current density.
\newblock \emph{Nature Communications}, 10\penalty0 (1), 2019.
\newblock \doi{10.1038/s41467-018-07792-9}.

\bibitem[Zhang and Zeng(2012)]{Zhang2012}
D.~Zhang and K.~Zeng.
\newblock Evaluating the behavior of electrolytic gas bubbles and their effect
  on the cell voltage in alkaline water electrolysis.
\newblock \emph{Industrial {\&} Engineering Chemistry Research}, 51\penalty0
  (42):\penalty0 13825--13832, 2012.
\newblock \doi{10.1021/ie301029e}.

\bibitem[Iwata et~al.(2021)Iwata, Zhang, Wilke, Gong, He, Gallant, and
  Wang]{Iwata2021}
R.~Iwata, L.~Zhang, K.~L. Wilke, S.~Gong, M.~He, B.~M. Gallant, and E.~N. Wang.
\newblock Bubble growth and departure modes on wettable/non-wettable porous
  foams in alkaline water splitting.
\newblock \emph{Joule}, 2021.
\newblock \doi{10.1016/j.joule.2021.02.015}.

\end{thebibliography}
\end{document}